\documentclass[12pt,journal,onecolumn]{IEEEtran}
\usepackage{latexsym}
\usepackage{epsf}
\usepackage{amsmath,amssymb,graphicx}
\usepackage{citesort}
\usepackage{float}
\newtheorem{lemma}{Lemma}[section]
\newtheorem{cor}{Corollary}[section]
\newtheorem{prop}{Proposition}[section]
\newtheorem{defn}{Definition}[section]

\newtheorem{claim}{Claim}[section]

\newcommand{\IR}{\mathbb{R}}
\newcommand{\IZ}{\mathbb{Z}}
\newcommand{\IC}{\mathbb{C}}
\newcommand{\IP}{\mathbb{P}}
\newcommand{\IN}{\mathbb{N}}
\newcommand{\ID}{\mathbb{D}}
\newcommand{\Cinfo}{C_{\rm {\it info}}}
\newcommand{\Cop}{C_{\rm {\it op}}}

\begin{document}

\title{Capacity per Unit Energy of Fading Channels with a Peak Constraint }
\author{Vignesh Sethuraman,{\it Member, IEEE}, and Bruce Hajek, {\it Fellow, IEEE}
\thanks{Manuscript received March 15, 2004; revised October 19, 2004; revised March 4, 2005. This work was supported in part by  NSF Grant ITR 00-85929, and by the Vodafone-US Foundation Graduate Fellowship.
The material in this paper was presented in part at IEEE ISIT, Yokohama, Japan, June 2003.  V. Sethuraman and B. Hajek are with the Department of Electrical and Computer Engineering and the Coordinated Science Laboratory, University of Illinois at Urbana-Champaign, Urbana, IL 61801 USA (email:\{vsethura,b-hajek\}@uiuc.edu). 
Communicated by Dr. Amos Lapidoth, Associate Editor for Shannon Theory.}}
\maketitle
\begin{abstract}
A discrete-time single-user scalar channel with temporally correlated Rayleigh fading is analyzed.  
There is no side information at the transmitter or the receiver.  
A simple expression is given for  the  capacity per unit energy, in the presence of a peak constraint.  
The simple formula  of Verd\'u  for capacity per unit cost is adapted to a channel with memory, and is used in the proof.  
In addition to bounding the capacity of a channel with correlated fading, the result gives some insight into the relationship between the correlation in the fading process and the channel capacity.  
The results are extended to a channel with side information, showing that the capacity per unit energy is one nat per Joule, independently of the peak power constraint. 

A continuous-time version of the model is also considered. The capacity per unit energy subject to a peak constraint (but no bandwidth constraint) is given by an expression similar to that for discrete time, and is evaluated for  Gauss-Markov and Clarke fading channels. 
\end{abstract}
\begin{center} \small\bfseries Index Terms \end{center}
Capacity per unit cost, channel capacity, correlated fading, flat fading, Gauss Markov fading
\vspace{0.67ex}

\section{INTRODUCTION}
Consider communication over a stationary Gaussian channel with Rayleigh flat fading.  The channel operates in  discrete-time, and there is no side information
about the channel at either the transmitter or the receiver. 
The broad goal is to find or bound the capacity of such a channel.
The approach taken is to consider the capacity per unit
energy.   Computation of capacity per unit energy is relatively tractable,
due to the simple formula of Verd\'u \cite{Verdu90} (also see Gallager
\cite{Gallager87}).   The study of capacity per unit energy naturally
leads one in the direction of low SNR, since capacity per unit energy
is typically achieved at low SNR.  
However, it is known that to achieve capacity or capacity per unit energy  
at low SNR, the optimal input signal becomes increasingly bursty 
\cite{AbouFaycalTrottShamai01,MedardGallager02,SubramanianHajek02}.
Moreover, such capacity per unit energy becomes the same as for the
additive Gaussian noise channel, and the correlation function of the
fading process does not enter into the capacity.
This is not wholly satisfactory, both because very large burstiness is often
not practical, and because one suspects that the correlation function
of the fading process is relevant.

To model the practical infeasibility of using large peak powers,  this paper
investigates the effect of hard-limiting the energy of each input
symbol by some value $P$.   A simple expression is given for the capacity per unit energy under such a peak constraint.
The correlation of the fading process enters into the capacity expression
found. 

When channel state information  is available  at the receiver (coherent channel), the capacity per unit energy under a peak constraint evaluates to one nat per Joule.
Continuous time channels are also considered. An analogous peak power constraint is imposed on the input signal. The capacity per unit energy expression is similar to that for the discrete-time channel.

An alternative approach to constraining input signal burstiness is to constrain the fourth moments, or kurtosis, of input signals \cite{MedardGallager02,SubramanianHajek02,Hassibi04}. 
This suggests evaluating the capacity per unit energy subject to a fourth moment constraint on the input. 
 We did not pursue the approach because it is not clear how to capture the constraint in the capacity per unit cost framework, whereas a peak constraint simply restricts the input alphabet. Also, a peak constraint is easy to understand, and matches well with popular modulation schemes such as phase modulation.  
Since a peak constraint $|X| \leq \sqrt{P}$ on a  random variable $X$  implies  $E[X^4] \leq P E[X^2]$, the bound of  M\'{e}dard and Gallager \cite{MedardGallager02} involving fourth moments yields a bound for a peak constraint, as detailed in Appendix \ref{app:FourthegyBoundDerivation}. 

The results offer some insight into the effect that correlation in the
fading process has on the channel capacity.  
There has been considerable progress on computation of capacity for
fading channels (see for example Telatar \cite{Telatar99}, and
Marzetta and Hochwald \cite{MarzettaHochwald99}).
This paper examines a channel with stationary temporally correlated Gaussian fading. 
The notion of capacity per unit energy is especially relevant for channels with low signal to noise ratio. 
Fading channel capacity for high SNR has recently been of interest (see \cite{Lapidoth05} and references therein).

The material presented in this paper is related to some of the material in \cite{Verdu02} and \cite{HajekSubramanian02}.
Similarities of this paper to \cite{Verdu02} are that both  consider the
low SNR regime, both have correlated fading, and the correlation
of the fading is relevant in the limiting analysis. An important difference is that
\cite{Verdu02} assumes the receiver knows the channel. Other differences
are that, here,  a peak constraint is imposed, the wideband
spectral efficiency is not considered, and the correlation is in time rather than
across antennas.  Similarities of this paper with  \cite{HajekSubramanian02}
are that both impose a peak constraint, but in \cite{HajekSubramanian02}
only the limit of vanishingly small peak constraints is considered, and
correlated fading random processes are not considered. 
The papers \cite{MedardGallager02} and \cite{SubramanianHajek02} are also related. They  are more general in that doubly-selective fading is considered, but they do not consider a peak constraint.

The organization of this paper is as follows. 
Preliminary material on capacity per unit time and per unit cost of fading channels with memory is presented in Section \ref{sec.Preliminaries}. 
The formula for capacity per unit energy for Rayleigh fading is presented in Section \ref{sec.MainResults}. 
The results are applied in Section \ref{sec.Examples} to two specific fading models, namely, the Gauss Markov fading channel, and the Clarke fading channel.
Proofs of the results are organized into Sections \ref{sec.PreliminaryProofs} -- \ref{sec.ExtensionContinuousTime}.
The conclusion is in Section \ref{sec.conclusion}.  All capacity computations are
in natural units for simplicity. One natural unit, nat, is
$\frac{1}{\log(2)}=1.4427$ bits.

\section{Preliminaries}  \label{sec.Preliminaries} 
Shannon \cite{Shannon49} initiated the study of information to cost ratios. 
For discrete-time memoryless channels without feedback, Verd\'{u} \cite{Verdu90} showed that, in the presence of a unique zero-cost symbol in the input alphabet, the capacity per unit cost is given by maximizing a ratio of a divergence expression to the cost function.  
The implications of a  unique zero-cost input symbol were studied by Gallager \cite{Gallager87} in the context of reliability functions per unit cost. 
In this section, the theory of capacity per unit cost is  adapted to fading channels with memory with the cost metric being transmitted energy. 
Additionally, a peak constraint is imposed on the input alphabet. 

Consider a single-user discrete-time channel without channel state information at either transmitter or receiver. The channel includes additive noise and multiplicative noise (flat fading), and  is specified by
\begin{equation}
Y(k)=H(k)X(k)+W(k), ~~k  \in Z \label{eq:ChannelModel}
\end{equation}
where $X$ is the input sequence, $H$ is the fading process, $W$ is an additive noise process, and $Y$ is the output. 
The desired bounds on the average and peak transmitted power are denoted by 
$P_{ave}$ and $P_{peak}$. 

An $(n,M,\nu, P, \epsilon)$ code for this channel consists of M codewords, each of block length $n$, such that each codeword ($X_{m1}$,\ldots,$X_{mn}$), $m=1,\ldots,M$, satisfies
the constraints
\begin{eqnarray} 
\sum_{i=1}^n |X_{mi}|^2&\leq \nu, \label{eq:AverageCostConstraint}\\
\max_{1\leq i \leq n}  |X_{mi}|^2 & \leq P, \label{eq:PeakCostConstraint}
\end{eqnarray}
and the average (assuming equiprobable messages) probability of
decoding the correct message is greater than or equal to  $1-\epsilon$.

Two definitions of capacity per unit time for the above channel are now considered. Their equivalence is established in Proposition \ref{prop.CopCinfo} for a certain class of channels. 
Capacity per unit energy is then defined and related to the definitions of capacity per unit time, and a version of Verd\'{u}'s formula is given. 
\begin{defn} {\em Operational capacity:} \label{defn:OperationalDefnCapacity}
A number $R$ is an $\epsilon$-achievable rate per unit time
if for every $\gamma > 0$, there exists $n_o$ sufficiently large so that if $n \geq n_o$,
there exists an $(n,M,nP_{ave}, P_{peak}, \epsilon)$ code with $\log M \geq n(R-\gamma)$.
A nonnegative number $R$ is an achievable rate per unit time if it is $\epsilon$-achievable 
for $0< \epsilon < 1$.   The operational capacity, $\Cop(P_{ave}, P_{peak})$, is the
maximum of the achievable rates per unit time.
\end{defn}

For any $n \in \IN$ and $P>0$, let
\begin{equation}
{\ID}_n(P) = \{ x\in \IC^n : |x_i|^2 \leq P \mbox{ for } 1\leq i \leq n\}. \label{eq:DefnOfG}
\end{equation}
\begin{defn} {\em Information theoretic capacity:} \label{defn:InfoTheoryDefnCapacity}
The mutual information theoretic capacity is defined as follows,
whenever the indicated limit exists:
\begin{equation}
\Cinfo (P_{ave}, P_{peak})=\lim_{n\rightarrow\infty} \sup_{P_{X_1^n}} \frac{1}{n}I(X_1^n;Y_1^n),
\end{equation}
where the supremum is over probability distributions $P_{X_1^n}$ on ${\ID}_n(P_{peak})$ such
that
\begin{equation}
\frac{1}{n}E_{P_{X_1^n}}[ ||X_1^n||_2^2]\leq P_{ave}.\label{eq:AvePowerConstraintDistribution}
\end{equation}
Similarly, $\overline{C}_{info}$ and $\underline{C}_{info}$ are defined by:
\begin{eqnarray}
\overline{C}_{info}(P_{ave}, P_{peak}) & = & \sup_n   \sup_{P_{X_1^n}} \frac{1}{n}I(X_1^n;Y_1^n), \\
\underline{C}_{info}(P_{ave}, P_{peak}) & = & \liminf_{n\rightarrow\infty}   \sup_{P_{X_1^n}} \frac{1}{n}I(X_1^n;Y_1^n),
\end{eqnarray}
where the suprema are over probability measures $P_{X_1^n}$ on ${\ID}_n(P_{peak})$ that satisfy (\ref{eq:AvePowerConstraintDistribution}).
\end{defn}
For memoryless channels, results in information theory imply  the equivalence of Definitions \ref{defn:OperationalDefnCapacity} and \ref{defn:InfoTheoryDefnCapacity}. 
This equivalence can be extended to channels with memory under mild conditions. 
In this regard, the following definitions for mixing, weakly mixing and ergodic processes are quoted from \cite[\S 5]{Maruyama49} for ease of reference (also see \cite[pp. 70]{Pinsker}). 

Let $\phi_i(z_1,z_2, \hdots, z_n)$ $(i=1,2)$ be bounded measurable functions of an arbitrary number of complex variables $z_1,\hdots,z_n$. 
Let $M_t$ be the operator ${\lim_{t\to \infty}\frac{1}{t}\sum_1^t}$ for discrete-time, and ${\lim_{t \to \infty}\frac{1}{t}\int_0^t dt}$ for continuous time.
A stationary stochastic process $z(t)$ ($t\in \IZ$ for discrete-time processes, and $t \in \IR$ for continuous-time processes\footnote{In this paper, continous-time processes are assumed to be mean square continuous.}) 
 is said to be:
\begin{enumerate}
\item strongly mixing (a.k.a mixing) if, for all choices of $\phi_1, \phi_2$, and times $t_1, \hdots t_n$,  $t_1^*,\hdots, t_n^*$, 
\begin{eqnarray}
\psi(t) &=& E[\phi_1(z(t_1), \hdots, z(t_n)) \cdot \phi_2(z(t_1^* + t),\hdots, z(t_n^*+t)]\nonumber\\
& & - E[\phi_1(z(t_1), \hdots, z(t_n))]\cdot E[\phi_2(z(t_1^*),\hdots, z(t_n^*)] \to 0 \mbox{ as } t \to \infty, \label{eq:MixingStronglyDefn}
\end{eqnarray}
\item weakly mixing if,  for all choices of $\phi_1, \phi_2$, and times $t_1, \hdots t_n$,  $t_1^*,\hdots, t_n^*$,  
\begin{equation}
M_t[\psi^2(t)] = 0, \label{eq:MixingWeaklyDefn}
\end{equation}
\item ergodic if,  for all choices of $\phi_1, \phi_2$, and times $t_1, \hdots t_n$,  $t_1^*,\hdots, t_n^*$, 
\begin{equation}
M_t[\psi(t)] = 0.
\end{equation}
\end{enumerate}
In general, strongly mixing implies weakly mixing, and weakly mixing implies ergodicity.  
Suppose a discrete-time or  continuous-time random process $H$ is a mean zero, stationary, proper complex Gaussian process.
Then, $H$ is weakly mixing if and only if its spectral distribution function $\{F_H(\omega):-\pi \leq \omega < \pi\}$ is continuous, or, equivalently, if and only if  
$ M_t[|R_H(t)|^2] =0$, where $R_H$ is the autocorrelation function of $H$. 
Also, $H$ is mixing if and only if $\lim_{t\rightarrow\infty} R_H(t)=0$  
\cite[Theorem 9]{Maruyama49}.  
It follows from the Riemann-Lebesgue theorem that H is mixing if $F_H$ 
is absolutely continuous. 
Furthermore, $H$ is ergodic if and only if it is weakly mixing.
To see this, it suffices to show that $H$ is not ergodic if $F_H$ is not continuous. 
Suppose $F_H$ has a discontinuity at, say,  $\lambda$. 
Let $
U_{\lambda} = M_t[H(t) ~e^{-i\lambda t}]
$. 
Clearly, $U_{\lambda}$ is zero-mean proper complex Gaussian. Also, $E[|U_{\lambda}|^2] = F_H(\lambda + 0) - F_H(\lambda - 0)$ \cite[Theorem 3]{Maruyama49}. 
Note that $| U_{\lambda} |$ is invariant to time-shifts of the process $H$. 
Since $|U_{\lambda}|$ is a non-degenerate shift-invariant function of $H$, it follows that $H$ is not an ergodic process \cite[5.2]{Pinsker}.  

The following proposition is derived from notions surrounding information stability (see \cite{Pinsker,Gray}) and the Shannon-McMillan-Breiman theorem for finite alphabet ergodic sources. 
A simple proof is given in Section \ref{sec.CopCinfoProof}. 
\begin{prop} \label{prop.CopCinfo}
If $H$ and $W$ are  stationary weakly mixing processes, and if $H,W$ and $X$ are
mutually independent, then for every $P_{ave}$, $P_{peak}>0$, $\Cinfo(P_{ave}, P_{peak})$
is well defined ($ \overline{C}_{info}(P_{ave}, P_{peak}) = \underline{C}_{info}(P_{ave}, P_{peak})$), and $\Cinfo(P_{ave}, P_{peak})=\Cop(P_{ave}, P_{peak})$. 
\end{prop}
Since ergodicity is equivalent to weakly mixing for Gaussian processes, 
the above proposition then implies that the two definitions of capacity coincide for the channel modeled in (\ref{eq:ChannelModel}) if $H$ and $W$ are stationary ergodic Gaussian processes and $H$, $W$ and $X$ are mutually independent. 

Following \cite{Verdu90}, the capacity per unit energy is defined along the lines of the operational definition of capacity per unit time, $C_{op}()$, as follows.
\begin{defn} 
Given $0\leq\epsilon <1$, a nonnegative number $R$ is an $\epsilon$-achievable
rate per unit energy with peak constraint $P_{peak}$ if for every $\gamma>0$, there exists $\nu_o$ large enough such
that if $\nu\geq\nu_o$, then an $(n,M,\nu, P_{peak}, \epsilon)$ code can be found with
$\log M > \nu (R-\gamma)$. A nonnegative number $R$ is an achievable rate per unit energy
if it is $\epsilon$-achievable for all $0<\epsilon<1$.   Finally, the {\em capacity $C_p(P_{peak})$ per unit energy} is the maximum achievable rate per unit energy.
\end{defn}
The subscript $p$ denotes the fact that a peak constraint is imposed.
It is clear from the definitions that, for any given $0< \epsilon <1$, if $R$ is an $\epsilon$-achievable rate per unit time, then $R/P_{avg}$ is an $\epsilon$-achievable rate per unit energy. 
It follows that $C_p(P_{peak})$ can be used to bound from above the capacity per unit time, $C_{op}(P_{ave}, P_{peak})$, for a specified peak constraint $P_{peak}$ and average constraint $P_{ave}$, as follows.
\begin{equation}
C_{op}(P_{ave},P_{peak}) \leq P_{ave}~ C_p(P_{peak}). \label{eq:CapacityUpperBoundUsingCapPerUnitEnergy}
\end{equation}
The following proposition and its proof are similar with minor differences to \cite[Theorem 2]{Verdu90}, given for memoryless sources.
\begin{prop}  \label{prop.peak_equiv}
 Suppose $C_{op}(P_{ave},P_{peak})=C_{info}(P_{ave},P_{peak})$ for $0 \leq P_{ave} \leq P_{peak}$ (see Proposition \ref{prop.CopCinfo} for sufficient conditions).
Then capacity per unit energy for a peak constraint $P_{peak}$ is given by 
\begin{eqnarray} \label{eq.cap}
C_p(P_{peak}) &=& \sup_{P_{ave} >  0} \frac{\Cop(P_{ave},P_{peak})}{P_{ave}} \label{eq:CapEnergyMutualInfoConnectVerdu0}\\
 &=&  \sup_n \sup_{ P_{X_1^n}}      \frac{ I(X_1^n; Y_1^n)}{ E[ \| X_1^n \|_2^2 ] } \label{eq:CapEnergyMutualInfoConnectVerdu}
\end{eqnarray}
where the last supremum is over probability distributions on ${\ID}_n(P_{peak})$. 
Furthermore, 
\begin{equation}
C_p(P_{peak})  =  \lim_{n\rightarrow \infty} \sup_{X \in {\ID}_n(P_{peak})} \frac{D(p_{Y|X}\|p_{Y|0})}{\|X\|_2^2}. \label{eq:CapacityDivergenceExpression}
\end{equation}
\end{prop}
The proof is given in Section \ref{sec.peak_equivProof}. 
For $P_{peak}$ fixed, $C_{op}(P_{ave}, P_{peak})$ is a concave non-decreasing function of $P_{ave}$. This follows from a simple time-sharing argument. 
It follows that 
\begin{equation}
\sup_{P_{ave} >  0} \frac{\Cop(P_{ave},P_{peak})}{P_{ave}} =  \lim_{P_{ave}\rightarrow 0} \frac{\Cop(P_{ave},P_{peak})}{P_{ave}}.
\end{equation}
So, the supremum in (\ref{eq:CapEnergyMutualInfoConnectVerdu0}) can be replaced by a limit. 

If $H$ and $W$ are i.i.d. random processes so that the channel is memoryless, then the suprema over $n$ in (\ref{eq:CapEnergyMutualInfoConnectVerdu}) and (\ref{eq:CapacityDivergenceExpression}) are achieved by $n=1$. 
Proposition \ref{prop.peak_equiv} then becomes a special case of Verd\'{u}'s results \cite{Verdu90}, which apply to memoryless channels with general alphabets and general cost functions. 

Equation (\ref{eq:CapacityDivergenceExpression}), which is analogous to \cite[Theorem 2]{Verdu90}, 
is especially useful because it involves a supremum over $\ID_n(P_{peak})$ rather than over probability distributions on $\ID_n(P_{peak})$.
This is an important benefit of considering capacity per unit cost when there is a zero cost input. 
It is noted that the natural extension of the corollary following \cite[Theorem 2]{Verdu90} also applies here. The proof is identical: 

\begin{cor}
Suppose $C_{op}(P_{ave},P_{peak})=C_{info}(P_{ave},P_{peak})$ for $0 \leq P_{ave} \leq P_{peak}$ (see Proposition \ref{prop.CopCinfo} for sufficient conditions).
Rate $R$ is achievable per unit energy with peak constraint $P_{peak}$ if and only if for every $0 < \epsilon < 1$ and $\gamma >0$, there exist $s>0$ and $\nu_0$, such that if $\nu \geq \nu_0$, then an $(n,M,\nu,P_{peak},\epsilon)$ code can be found with $\log M > \nu (R-\gamma)$ and $n<s\nu$. 
\end{cor}

For the remainder of this paper, the fading process $H$ is assumed to be stationary and ergodic. 
Both $H$ and the additive noise $W$ are modeled as zero mean proper complex Gaussian processes, and without loss of generality, are normalized to have unit variance. 
Further, $W$ is assumed to be a white noise process. 
The conditions of Proposition \ref{prop.CopCinfo}  are satisfied, and so the two definitions of capacity per unit time are equivalent. 
Henceforth, the capacity per unit time is denoted by $C(P_{ave}, P_{peak})$. 
Also, for brevity, in the remainder of the paper, a peak power constraint is often denoted by $P$ instead of $P_{peak}$. 

\section{Main results} \label{sec.MainResults}
\subsection{Discrete-time channels}
The main result of the paper is the following.
\begin{prop} \label{prop.main}
Let $S(\omega)$ denote the density of the absolutely continuous component of the power spectral measure of $H$.
The capacity per unit energy for a finite peak constraint $P$ is
given by 
\begin{eqnarray}
C_p(P) & = &1-\frac{I(P)}{P}, \label{eq:EnergyBound}\\
\mbox{where } I(P) & = & \int_{-\pi}^{\pi} \log(1+P S(\omega))\,\frac{d\omega}{2\pi}. \label{eq:mainprop_I}
\end{eqnarray}
\end{prop}
Moreover, roughly speaking, the capacity per unit energy $C_p(P)$ can be asymptotically achieved using codes with the following structure. 
Each codeword is ON-OFF with ON value $\sqrt{P}$. 
The vast majority of codeword symbols are OFF, with infrequent long bursts of ON symbols. See the end of Section \ref{subsec.OptimalityOfTemporalOnOffSignaling} for a more precise explanation.

Suppose that in the above channel model, channel side information (CSI) is available at the receiver. The fading process is assumed to be known causally at the receiver; i.e. at time step $k$, the receiver knows $\{ H(n): n\leq k \}$.   
For this channel, a $(n,M,\nu, P, \epsilon)$ code, achievable rates and the capacity per unit energy for peak constraint $P$, denoted by $C_p^{coh}(P)$, are respectively defined in a similar manner as for the same  channel without CSI.
\begin{prop} \label{prop.CSImain}
For $P > 0$, 
$C_p^{coh}(P) = 1$.
\end{prop}

There is an intuitively pleasing interpretation of Proposition \ref{prop.main}. 
Note that $C_p(P) = C_p^{coh}(P) -\frac{1}{P} I(P)$. The term $\frac{1}{P}I(P)$ can be interpreted as the penalty for not knowing the channel at the receiver. The integral $I(P)$ is the information rate between the fading channel process and the output when the signal is deterministic and identically $\sqrt{P}$ (see Section  \ref{subsec.IdentifyingTheLimit}). 
When ON-OFF signaling is used with ON value $\sqrt{P}$ and long ON times, the receiver gets information about the fading channel at rate $I(P)$ during the ON periods, which thus subtracts from the information that it can learn about the input. 
Similar observations have been previously made  in different contexts \cite{Biglieri98,Jacobs63}. 

The definition of $C_p(P)$ still makes sense if $P=\infty$, and $C_p(\infty)$ is the capacity per unit energy with no peak constraint. It is well known that $C_p(\infty) = 1$ (see  \cite[p. 816]{SubramanianHajek02}, \cite{Jacobs63,Kennedy69,TelatarTse00}). Note that, by (\ref{eq:EnergyBound}) and (\ref{eq:mainprop_I}), as $P\to \infty$, $C_p(P) \to 1 = C_p(\infty)$. 
By their definitions, both $C_p^{coh}(P)$ and $C_p(\infty)$ are upper bounds for $C_p(P)$. The bounds happen to be equal: $C_p^{coh}(P) = C_p(\infty) = 1$.
				
Another upper bound on $C_p(P)$ is $U_p(P)$, defined by 
\begin{equation}
U_p(P)  =   \frac{P}{2}\int_{-\pi}^{\pi}S^2(\omega)\,\frac{d\omega}{2\pi}. \label{eq:FourthegyBound}
\end{equation}
The fact $C_p(P)\leq U_p(P)$ follows easily from (\ref{eq:EnergyBound}), (\ref{eq:mainprop_I}) and the inequality $\log(1+x) \geq x-\frac{x^2}{2}$ for $x\geq0$. 
Also, $C_p(P) \to U_p(P)$ as $P \to 0$. It is shown in Appendix \ref{app:FourthegyBoundDerivation} that the bound $U_p(P)$ is also obtained by applying an inequality of M\'{e}dard and Gallager \cite{MedardGallager02}.

\subsection{Extension to continuous-time channels}
The model for continuous time is the following.   
Let $(H(t): -\infty < t <\infty)$ be a continuous-time stationary ergodic  proper complex Gaussian process such that $E[|H(t)|^2]=1$. 
A codeword for the channel is a deterministic signal $X=(X(t):0\leq t \leq T)$, where $T$ is the duration of the signal.  
The observed signal is given by
\begin{equation}
Y(t)=H(t)X(t)+W(t), ~0 \leq t \leq T \label{eq:ChannelModelContinuous}
\end{equation}
where $W(t)$ is a complex proper Gaussian white noise process with $E[W(s)\overline{W(t)}]=\delta(s-t)$.
The mathematical interpretation of this, because of the white noise, is that the integral process  
$V=(V_t=\int_0^t Y(s) ds: 0 \leq t \leq T )$ is observed  \cite{SubramanianHajek02}.  
The mathematical model for the observation process is then 
\begin{equation}
V(t)=\int_0^t H(s) X(s)  ds + \eta(t) ~~~0 \leq t \leq T,  \label{eq:Vrepresentation}
\end{equation}
where $\eta$ is a standard proper complex Wiener process with autocorrelation function $E[\eta(s)\overline{\eta(t)}]=\min\{s,t\}$.
The process $V$ takes values in the space of continuous functions on $[0,T]$ with $V(0)=0$.

A $(T,M,\nu, P,\epsilon)$ code for the continuous-time channel is  defined analogously to an $(n,M,\nu,P,\epsilon)$ code for the discrete-time channels, with the block length $n$ replaced by the code duration $T$, and the constraints (\ref{eq:AverageCostConstraint}) and (\ref{eq:PeakCostConstraint})  replaced by
\begin{eqnarray}
\int_0^T |X(t)| ^2  dt & \leq & \nu, \\
\sup_{0\leq t \leq T}  |X(t)|^2  & \leq & P.
\end{eqnarray}
The codewords are required to be Borel measurable functions of $t$, but otherwise, no bandwidth restriction is imposed. 
Achievable rates and the capacity per unit energy for peak constraint $P$, denoted $C_p(P)$, are defined as for the discrete-time channel.

\begin{prop}  \label{prop.continuous}
Let $S_H(\omega)$ denote  the density of the absolutely continuous component of  the power spectral measure of $H$. Then 
\begin{eqnarray}
C_p(P) &=& 1-\frac{I(P)}{P}, \label{eq:PropContinuousMainEquation} \\
\mbox{where } I(P) &=& \int_{-\infty}^\infty \log(1+PS_H(\omega)) \frac{d\omega}{2\pi}. \label{eq:mainprop_Icontinuous}
\end{eqnarray}
\end{prop}
The proof is given in Section \ref{sec.ExtensionContinuousTime}.

The following upper bound $U_p(P)$ on $C_p(P)$ is constructed on the lines of the upper bound on the discrete-time capacity per unit energy defined in (\ref{eq:FourthegyBound}).
\begin{equation}
U_p(P)  =   \frac{P}{2}\int_{-\infty}^{\infty}S^2(\omega)\,\frac{d\omega}{2\pi} \label{eq:FourthegyBoundContinuous}
\end{equation}
Similar to the discrete-time case, $C_p(P) \to U_p(P)$ as $P \to 0$.

\section{ILLUSTRATIVE EXAMPLES } \label{sec.Examples}
Using Propositions \ref{prop.main} and \ref{prop.continuous}, the capacity per unit energy with a peak constraint is obtained in closed form for two specific models of the channel fading process. 
The channel models considered are Gauss-Markov fading and Clarke's fading. 
Finally, the capacity per unit energy with peak constraint is evaluated for a block fading channel with constant fading within blocks and independent fading across blocks.

\subsection{ Gauss-Markov Fading} \label{sec.GaussMarkov}
\subsubsection{Discrete-time channel}
Consider the channel modeled in (\ref{eq:ChannelModel}). Let the fading process $H$ be Gauss-Markov with autocorrelation function $R_H(k)=\rho^{|k|}$ for some $\rho$ with $0 \leq  \rho < 1$. 
\begin{cor} \label{cor.GaussMarkov}
The capacity per unit energy for peak constraint $P$, for the Gauss-Markov fading channel, is given by 
\begin{equation}
C_p(P)=1-\frac{\log(z_+)}{P} \label{eq:EnergyBoundGaussMarkov}
\end{equation}
 where $z_+$ is the larger root of the quadratic equation $z^2-(1+P+\rho^2(1-P))z+\rho^2 =0$.  The bound $U_p(P)$ simplifies to the following:
\begin{equation}
U_p(P) = \frac{P}{2}\left( \frac{1+\rho^2}{1-\rho^2} \right)  \label{eq:FourthegyBoundGaussMarkov}
\end{equation}
\end{cor}
For the proof of the above corollary, see Appendix  \ref{app.GaussMarkovProof}.

\begin{figure}
\begin{center}
\epsfbox{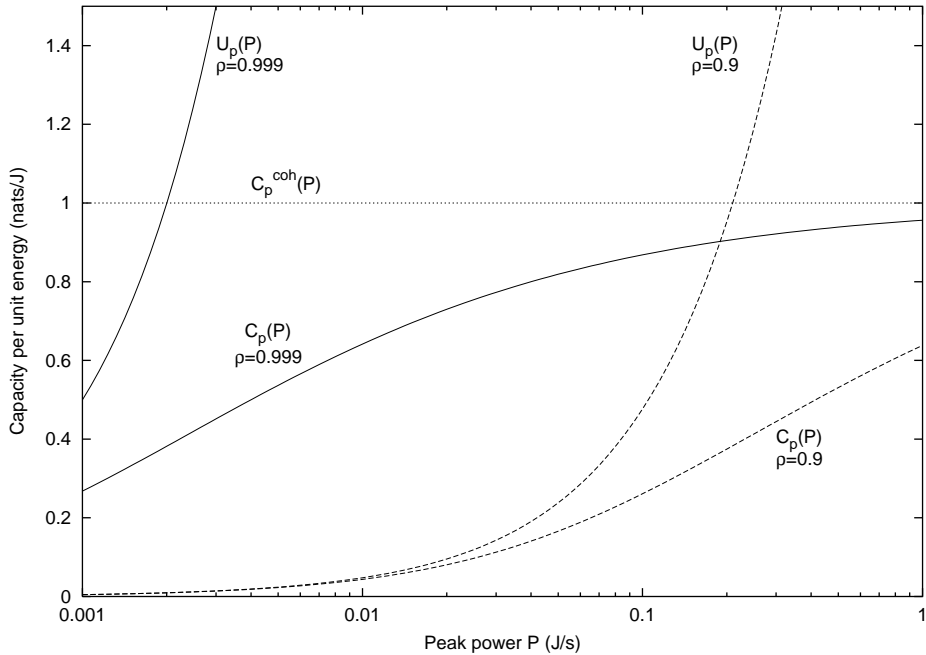} 
\caption{Capacity per unit energy, and upper bounds for the discrete-time channel, for $\rho=0.9$, $\rho=0.999$} \label{fig:CapPerUnitEnergyDiscreteBounds}
\end{center}
\end{figure}

The upper bounds $C_p^{coh}(P)$ and $U_p(P)$ are compared to $C_p(P)$ as a function of peak power $P$ in Figure \ref{fig:CapPerUnitEnergyDiscreteBounds} for $\rho =0.9$ and $\rho = 0.999$. 
\begin{figure}
\begin{center}
\epsfbox{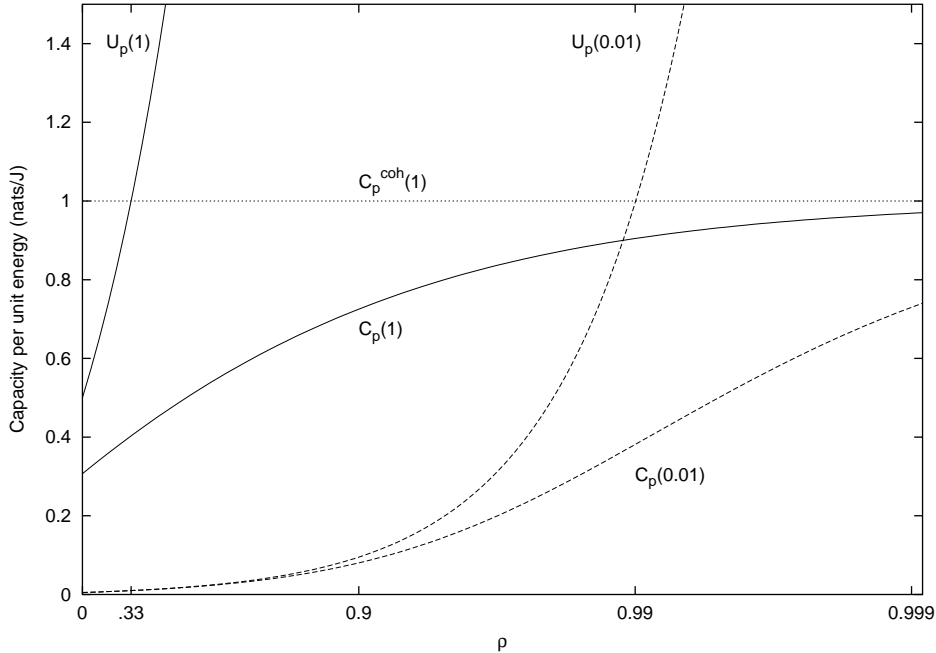}
\caption{Capacity per unit energy as a function of $\rho$, for the discrete-time channel}  \label{fig:BoundsRhoDiscrete}
\end{center}
\end{figure}
The figures illustrate the facts that  $C_p(P) \to C_p^{coh}(P)$ in the limit as $P \to \infty$, i.e., when the peak power constraint is relaxed, and that $C_p(P) \simeq U_p(P)$ as $P \to 0$.
In Figure \ref{fig:BoundsRhoDiscrete}, $C_p(P)$ and $U_p(P)$ are plotted as functions of the $\rho$, for various values of peak constraint $P$.

It is common in some applications to express  the peak power constraint as a multiple of the average power constraint. Consider   such a relation, where the peak-to-average ratio is constrained by a constant $\beta$, so
$$P = \beta P_{avg}.$$
From (\ref{eq:CapacityUpperBoundUsingCapPerUnitEnergy}) and (\ref{eq:FourthegyBound}), we get the following bounds on the channel capacity per unit time.  
To get the final expressions in (\ref{eq:EnergyBoundGaussMarkovCapPerUnitTime}) and (\ref{eq:FourthegyBoundGaussMarkovCapPerUnitTime}), $P$ is substituted by $\beta P_{avg}$ in the expressions for $C_p(P)$ and $U_p(P)$ in (\ref{eq:EnergyBoundGaussMarkov}) and  (\ref{eq:FourthegyBoundGaussMarkov}).
\begin{eqnarray}
C(\beta P_{avg}, P_{avg}) & \leq & C_p^{coh}(\beta P_{avg}) \cdot P_{avg} = 1 \cdot P_{avg} \\ 
C(\beta P_{avg}, P_{avg}) & \leq & C_p(\beta P_{avg}) \cdot P_{avg} = P_{avg} - \frac{\log(z_+^*)}{\beta} \label{eq:EnergyBoundGaussMarkovCapPerUnitTime} \\
C(\beta P_{avg}, P_{avg}) & \leq & U_p(\beta P_{avg}) \cdot P_{avg} = \left( \frac{1}{2} \cdot \frac{1+\rho^2}{1-\rho^2} \right) \beta P_{avg}^2 \label{eq:FourthegyBoundGaussMarkovCapPerUnitTime}
\end{eqnarray}
Here, $z_+^*$ is the larger root of the quadratic equation
$z^2-(1+\beta \cdot P_{avg}+\rho^2(1-\beta \cdot P_{avg}))z+\rho^2 =0$.

\begin{figure}
\begin{center}
\epsfbox{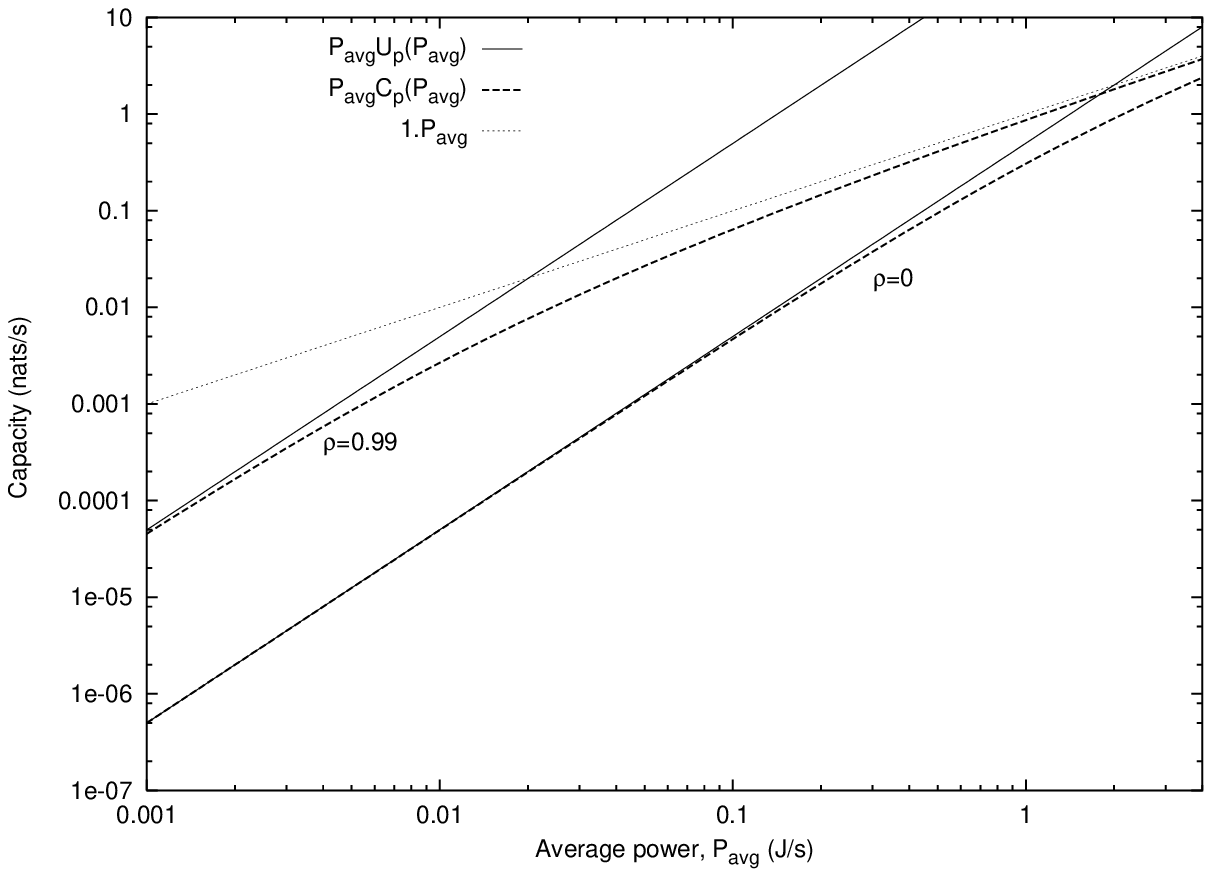}
\caption{Capacity bounds for the discrete-time channel, $\beta=1$} \label{fig:beta_1}
\epsfbox{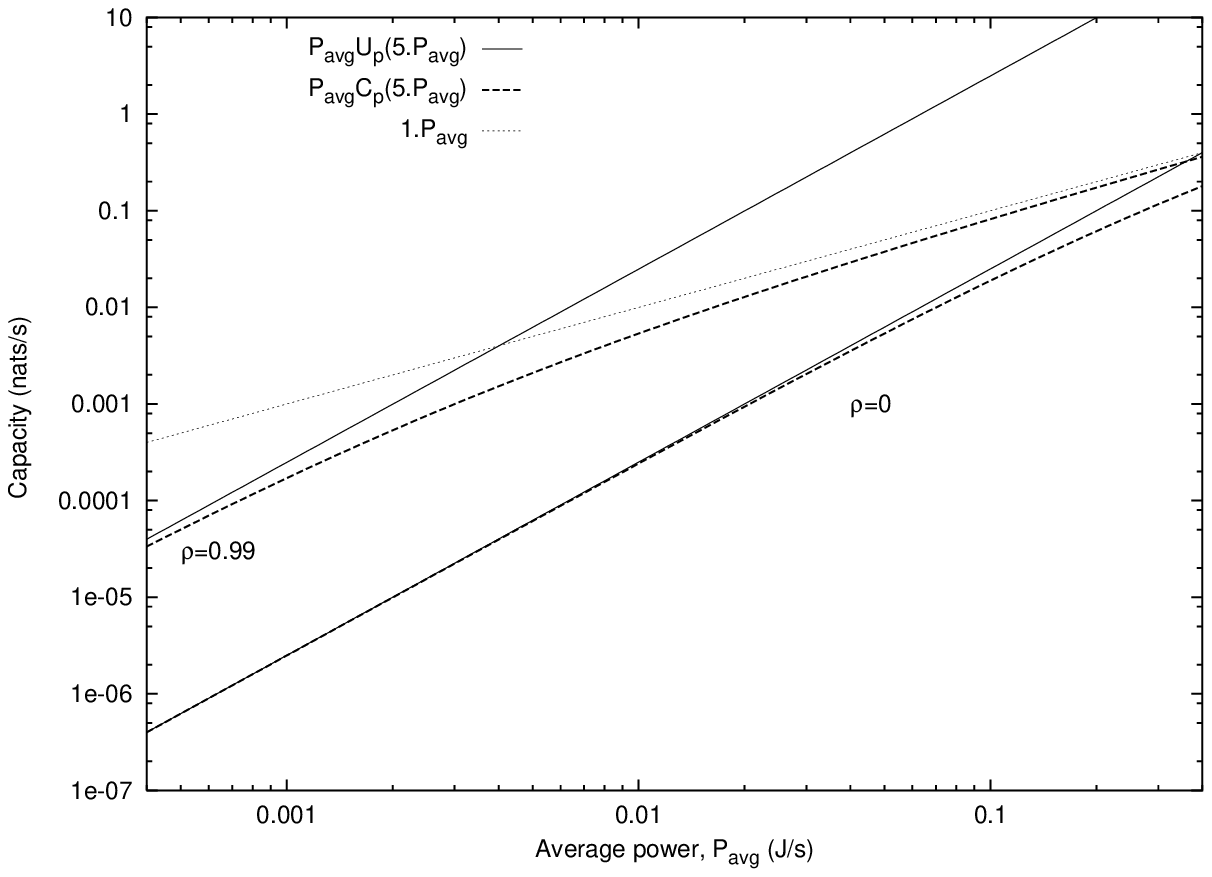}
\caption{Capacity bounds for the discrete-time channel, $\beta=5$} \label{fig:beta_5}
\end{center}
\end{figure}

The bounds are plotted for various values of $\rho$ and $\beta$ in Figures \ref{fig:beta_1} - \ref{fig:beta_5}. 
The average power $P_{avg}$ ($x$ axis) is in log scale.   All the capacity bounds converge at low power to zero. The fourthegy bound $U_p(P)$ tends to increase faster than $C_p(P)$ for higher $\beta$, i.e., more relaxed peak to average ratio. A similar behavior is observed when the  correlation coefficient $\rho$, and hence coherence time, is increased.
Note that the case when $\beta = 1$ corresponds to having only a peak power constraint, and no average power constraint.

\subsubsection{Continuous-time channel}
\begin{figure}
\begin{center}
\epsfbox{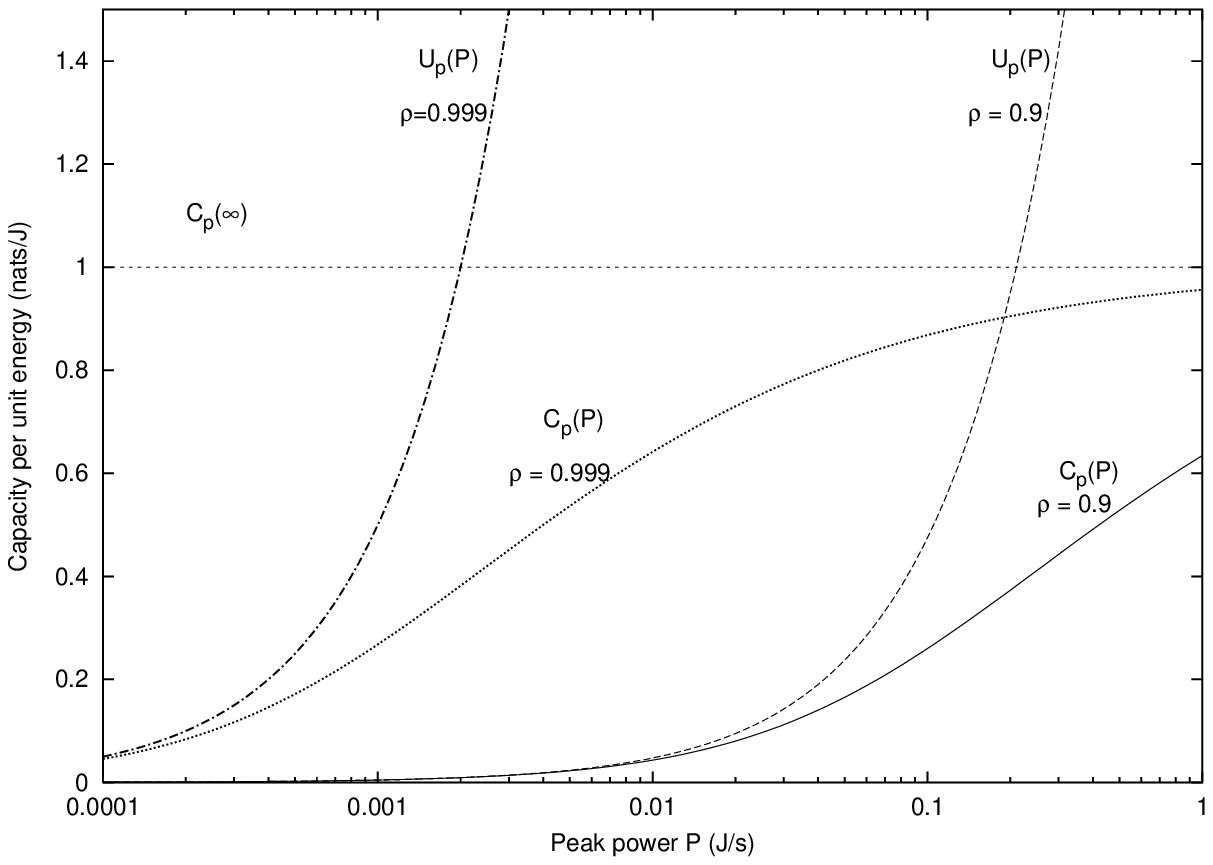}
\caption{Capacity per unit energy and upper bounds for the continuous-time channel, $\rho = 0.9$ and $\rho = 0.999$} \label{fig:CapPerUnitEnergyCtsBounds}
\epsfbox{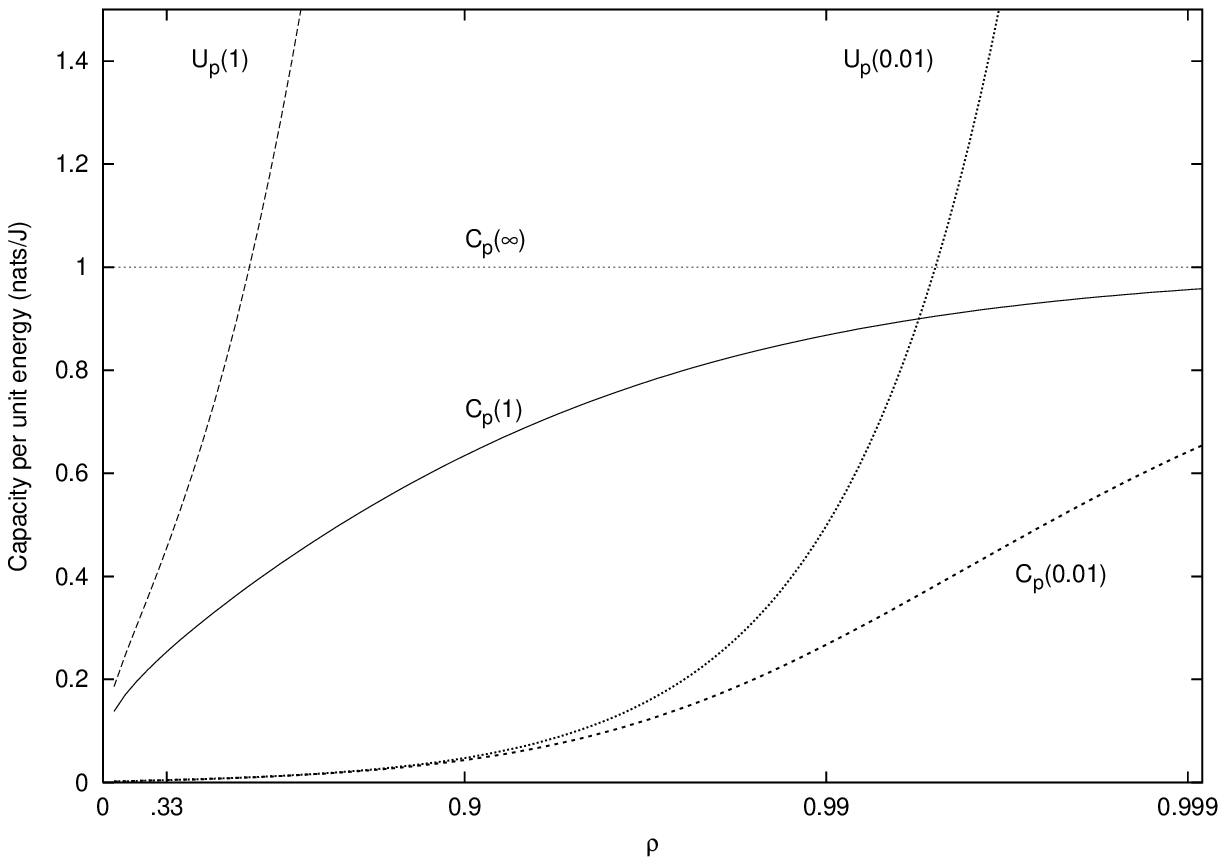}
\caption{Capacity per unit energy and upper bounds  as a function of $\rho$} \label{fig:BoundsRhoCts}
\end{center}
\end{figure}

Consider the channel modeled in (\ref{eq:ChannelModelContinuous}). As in the discrete-time case considered above, let the fading process $H$ be a Gauss-Markov process with autocorrelation $R_H(t)=\rho^{|t|}$, where the parameter $\rho$ satisfies $0 \leq  \rho < 1$.
The power spectral density $\{S(\omega): ~0\leq \omega \leq 2 \pi\}$ is given by 
\begin{equation}
S(\omega) = \frac{-2\log \rho}{\omega^2 + (\log\rho)^2}.
\end{equation}
The capacity per unit energy with peak constraint $P$ is obtained by using the above expression for the power spectral density in (\ref{eq:PropContinuousMainEquation}) and simplifying using the following standard integral \cite[Section 4.22, p. 525]{TableOfIntegralsGradshteyn}:
\begin{equation*}
\int_0^{\infty} \log \left(\frac{a^2+x^2}{b^2+x^2}\right) dx = (a-b) \pi, ~a>0, b>0
\end{equation*}
It follows that
\begin{equation}
C_p(P) = 1 - \frac 1 P \left( \sqrt{(\log \rho)^2 - 2P\log \rho} + \log \rho \right)
\end{equation}
The upper bound $U_p(P)$ in (\ref{eq:FourthegyBoundContinuous}) is evaluated using Parseval's theorem.
\begin{equation}
U_p(P) = \frac P {-2\log \rho}
\end{equation}
In Figure \ref{fig:CapPerUnitEnergyCtsBounds}, the capacity per unit energy with peak constraint $C_p(P)$ and the  upper bound $U_p(P)$ are plotted and compared as functions of peak power $P$ for various values of $\rho$. In Figure \ref{fig:BoundsRhoCts}, $C_p(P)$ and $U_p(P)$ are plotted as functions of the $\rho$, for various values of peak constraint $P$.

\subsection{Clarke's Fading} \label{sec.Clarke}
Fast fading manifests itself as rapid variations of the received signal envelope as the mobile receiver moves through a field of local scatterers (in a mobile radio scenario). 
Clarke's fading process \cite[Chapter 2 p.41]{Stuber} is a continuous-time proper complex Gaussian process with power spectral density  given by 
\begin{equation}
S(2 \pi f) = \left\{ \begin{array}{cc}
		\frac{1}{\pi f_m} \frac{1}{\sqrt{1 - (f/f_m)^2}} & |f| < f_m \\
		0	 & \mbox{elsewhere} \end{array} \right. \label{eq:DopplerSpectrum}
\end{equation}
where $f_m$ is the maximum Doppler frequency shift and is directly proportional to the vehicle speed. The model is based on the assumption of isotropic local scattering in two dimensions.
Consider a continuous-time channel modeled in (\ref{eq:ChannelModelContinuous}), with the fading process following the Clarke's fading model.
\begin{cor} \label{cor.time_selective}
For a time-selective fading process $H$ with the power spectral density given by (\ref{eq:DopplerSpectrum}), the capacity per unit energy for peak constraint $P$ is given by $C_p(P) =   g(P)- \frac{1}{P}\log( \frac{P}{2} ) +1 - \frac{\pi}{2}$, where $g(P)$ is given by
\begin{equation*}
g(P) = \left\{ \begin{array}{cc}
			 \sqrt{1-P^{-2}} \left( \frac{\pi}{2}-\arctan{\frac{1}{\sqrt{P^2-1}}} \right)  & P \geq 1 \\
			 \sqrt{P^{-2}-1} \left( Im( \arctan{\frac{1}{\sqrt{P^2-1}} }) \right) & P < 1 \end{array} \right. 
\end{equation*}
Here, $Im(z)$ is the imaginary part of the  complex number $z$.
\end{cor}
Corollary \ref{cor.time_selective} is obtained by evaluating the integral in (\ref{eq:mainprop_Icontinuous}). 
Note that, for this channel, the integral in (\ref{eq:FourthegyBound}) diverges, so that $U_p(P) \equiv +\infty$.

\subsection{Block Fading}
Suppose the channel fading process $H$, in either discrete-time (\ref{eq:ChannelModel}) or continuous-time (\ref{eq:ChannelModelContinuous}), is replaced by 
a block fading process with the same marginal distribution, but which is constant within each block (of length $T$) and independent across blocks.  
\begin{cor} \label{cor.BlockFadingDiscreteAndCts}
The capacity per unit energy with peak constraint $P$ of a block fading channel (discrete-time or continuous-time), block length $T$,  is  given by $C_p^B(P,T)=1-\frac{\log(1+PT)}{PT}$. 
\end{cor}
See Appendix \ref{app:BlockFadingProof} for the proof.
Note that for $P$ fixed, $\lim_{T\rightarrow \infty} C_p^B(P,T)=1$. 

The capacity per unit time of the above channel with peak constraint $P$ and average power constraint $P_{avg}$, denoted by $C_p^{B,T}(P, P_{avg})$, can be bounded from above using Corollary \ref{cor.BlockFadingDiscreteAndCts}, (\ref{eq:CapacityUpperBoundUsingCapPerUnitEnergy}) and the inequality $\log(1+x) \geq x-x^2/2$ for  $x\geq 0$ as follows. 
\begin{equation}
C_p^{B,T}(P, P_{avg}) \leq \frac T 2 P_{avg}\cdot P 
\end{equation}
In the presence of a peak-to-average ratio constraint, $\beta$ say, the bound on the capacity is quadratic in $P_{avg}$ for small values of $P_{avg}$. 
Similarly, the mutual information in  a Rayleigh fading channel (MIMO setting) is shown  in \cite{Hassibi04} to be quadratic in $P_{avg}$, as $P_{avg} \to 0$. 
\section{Proofs of Propositions in Section \ref{sec.Preliminaries}} \label{sec.PreliminaryProofs}

\subsection{Proof of Proposition \ref{prop.CopCinfo}} \label{sec.CopCinfoProof}
It is first proved that $\underline{C}_{info} \geq \Cop$.  Given $\epsilon > 0$, for all
large $n$ there exists an $(n,M,nP_{ave}, P_{peak},\epsilon)$ code with $\frac{1}{n}\log M \geq \Cop
-\epsilon$.  Letting $X_1^n$ represent a random codeword, with all $M$ possibilities having
equal probability,  Fano's inequality implies that $I(X_1^n;Y_1^n) \geq (1-\epsilon)\log M - \log 2$,
so that $I(X_1^n;Y_1^n) \geq (1-\epsilon)n(\Cop-\epsilon)  - \log 2.$    Therefore,
$\underline{C}_{info} \geq (1-\epsilon)(\Cop - \epsilon)$.   Since $\epsilon$ is arbitrary, the
desired conclusion,   $\underline{C}_{info} \geq \Cop$, follows.  It remains to prove the reverse inequality.
  
Consider the definition of $\Cinfo$. For $P_{peak}$ fixed, using a simple time-sharing argument, it can be shown that $\overline{C}_{info}(P_{ave}, P_{peak})$ is a concave non-decreasing function of $P_{ave}$.
Consequently,
given $\epsilon >0$, there exists an $n\geq 1$,  $\delta > 0$, and distribution $P_{X_1^n}$ on $\ID_n(P_{peak})$ (\ref{eq:DefnOfG}) such that $\frac{1}{n} I(X_1^n;Y_1^n) \geq \overline{C}_{info} - \epsilon$
with $ E[||X_1^n||^2]\leq nP_{ave}-2\delta$. 

Since the mutual information between
arbitrary random variables is the supremum of the mutual information between quantized
versions of the random variables \cite[2.1]{Pinsker},  there exist vector quantizers
$q:{\ID}_n(P_{peak})\rightarrow A$ and $r: \IC^n \rightarrow B$,  where $A$ and $B$ are finite sets, such
that $\frac{1}{n}I(q(X_1^n); r(Y_1^n))\geq \frac{1}{n}I(X_1^n; Y_1^n)-\epsilon \geq
 \overline{C}_{info} - 2\epsilon$.
By enlarging $A$ if necessary,  it can be assumed that for each $a$, $q^{-1}(a)$ is a subset of
one of the energy shells $\{ x\in {\ID}_n(P_{peak}):  K_a\delta \leq ||x||^2_2 \leq (K_a+1)\delta \}$ for some integer $K_a$.   Therefore, with $l$ defined on $A$ by $l(a)=\sup\{||x||^2_2: x\in q^{-1}(a) \}$, it follows that
$||x||_2^2 \geq l(a)-\delta$ for all $x\in q^{-1}(a)$, for all $a\in A$.   Hence,
\begin{equation} \label{eq.qproperty}
E[l(q(X_1^n))] \leq E[||X_1^n||_2^2]+\delta \leq nP_{ave}- \delta.
\end{equation}

For each $a\in A$, let $\gamma_a$ be the conditional probability measure of $X_1^n$ given
that $q(X_1^n)=a$.   Then $\gamma=(\gamma_a : a\in A)$ is the transition kernel for a memoryless
channel with input alphabet $A$ and output alphabet ${\ID}_n(P_{peak})$.   Define a new channel
$\tilde{\nu}$, with input alphabet $A$ and output alphabet $B$, as the concatenation of three
channels: the memoryless channel with transition kernel $\gamma$, followed by the original fading
channel, followed by the deterministic channel given by the quantizer $r$.  The idea of the remainder of the proof is that codes for channel $\tilde{\nu}$ correspond to random codes for the
original channel.

Let $(\hat{X}_k: k\in \IZ)$ consist of independent random variables in $A$, each with the distribution of
$q(X_1^n)$.  Let $(\hat{Y}_k : k\in \IZ)$ be the corresponding output of $\tilde{\nu}$
in $B^{\IZ}$.  Note that
$(\hat{X}_1, \hat{Y}_1)$ has the same distribution as $(q(X_1^n), r(Y_1^n))$, so that
$I(\hat{X}_1; \hat{Y}_1) = I(q(X_1^n); r(Y_1^n)) \geq n(\Cinfo - 2\epsilon).$
Since the input sequence $\hat{X}$ is i.i.d. and the channel is stationary,
$$
\frac{1}{k} I(\hat{X}_1^k; \hat{Y}_1^k)  \geq  I(\hat{X}_1; \hat{Y}_1) \geq n(\Cinfo-2\epsilon).
$$

Letting $(\hat{X}_k: k\in \IZ)$ be the input to a discrete memoryless channel with transition kernel $\gamma$ produces a process $X$ with independent length $n$ blocks, which can be arranged to form an i.i.d. vector process $(X_{(k-1)n + 1}^{kn}: k\in \IZ)$. 
Similarly, the processes $H$ and $W$ in the fading channel model can be arranged into blocks to yield vector processes.

It is now shown that for any discrete-time weakly mixing process $U$, the corresponding vector process obtained by arranging $U$ into blocks of length $n$ is also weakly mixing. 
For any choice of bounded measurable functions $\Phi_i$ ($i=1,2)$ on the vector process $(U_{(k-1)n+1}^{kn}: k \in \IZ)$, there exist corresponding functions $\phi_i$ defined on the process $U$. 
Let $\psi(t)$ be defined on $\phi_i$ as given in (\ref{eq:MixingStronglyDefn}), and $\Psi(t)$ be defined analogously on $\Phi_i$. 
Clearly, $\Psi(t) = \psi(nt)$ and 
\begin{eqnarray}
\frac 1 t \sum_1^t \Psi^2(i) &=& \frac 1 t \sum_1^t \psi^2(ni)\\
&\leq& n \frac 1 {nt} \sum_1^{nt} \psi^2 (i).
\end{eqnarray}
It follows that 
\begin{eqnarray}
M_t[\Psi^2(t)] & \leq & n M_t[\psi^2(t)] \\ 
	& = & 0 . \label{eq:WeaklyMixingImpliesBlockWeaklyMixing}
\end{eqnarray}
where (\ref{eq:WeaklyMixingImpliesBlockWeaklyMixing}) follows from the weakly mixing property of $U$ (\ref{eq:MixingWeaklyDefn}). Consequently, the vector process obtained from $U$ is also weakly mixing. 
Since $H$ and $W$ are weakly mixing, it follows that the corresponding vector processes are also weakly mixing. 

{\bf Remark:} It should be noted that, when an ergodic discrete-time process is arranged in blocks to form a vector process, the resulting vector process is not necessarily ergodic. 
For example, consider the following process  $U$. 
Let $U_0$ be $0$ or $1$ with equal probability. 
Let $U_1 = 1-U_0$, 
and $U_k = U_{k\mod 2}$. The process $U$ is ergodic. However, when $U$ is  arranged into a vector process of length $n=2$ (or any other even $n$), the vector process  is not ergodic. 
In fact, there exist ergodic processes such that the derived vector processes of length $n$ are not ergodic for any $n>1$. 
For example, let $U_k = \sum_{i\in \IP} V_k^{(i)}/2^i$, where the processes $V^{(i)}$ are independent, and for each process  $V^{(i)}$, $(V_1^{(i)}, \hdots,V_i^{(i)})$ is  chosen to be one of the $i$ patterns $(0\hdots 0,1)$, $(0\hdots0,1,0)$, $\hdots (1,0 \hdots0)$ with probability $1/i$, and $V_k^{(i)} = V_{k\mod i}^{(i)}$. 
Here, $\IP$ is the set of primes. 
It can be shown that $U$ is ergodic. 
For any $n \in \IP$ and $k\in \IN$, when $V^{(n)}$ is arranged into a vector process of length $kn$, the vector process is not ergodic. 
Since any $m \in \IN$ has factors in $\IP$, it follows that the vector process of length $m$ obtained from $U$ is not ergodic either. 

Arranging the output process $Y$ of the fading channel into a length-$n$ vector process, it is clear that the $k^{th}$ element of this process $Y_{(k-1)n+1}^{kn}$  depends only on the $k^{th}$ elements of the vector processes of $X$, $H$ and $W$. Further, the output $\hat{Y}_k$ is a function of $Y_{(k-1)n+1}^{kn}$. 
Therefore, the process 
($\hat{X}_k$,  $X_{(k-1)n+1}^{kn}$, 
 $H_{(k-1)n+1}^{kn}$, $W_{(k-1)n+1}^{kn}$, 
 $Y_{(k-1)n+1}^{kn}$, $\hat{Y}_k: k \in \IZ$) 
is a weakly mixing process. 
So, $\hat{X}$ and $\hat{Y}$ are jointly weakly mixing and hence jointly ergodic. 

Thus, the following limit exists:  $\overline{I}(\hat{X},\hat{Y}) = \lim_{k\rightarrow\infty}
\frac{1}{k}  I(\hat{X}_1^k  ,\hat{Y}_1^k) $, and the limit satisfies
$ \overline{I}(\hat{X},\hat{Y})  \geq  n(\Cinfo-2\epsilon)$. 
Furthermore, the asymptotic equipartition property (AEP, or Shannon-McMillan-Brieman theorem) for finite alphabet sources implies that
\begin{equation} 
\lim_{k\rightarrow \infty} P[|\frac{1}{k} i_k(\hat{X}_1^k;\hat{Y}_1^k)-\overline{I}(\hat{X};\hat{Y})|>\epsilon]=0, \label{eq:AEPFiniteAlphabet}
\end{equation} 
where $ i_k(\hat{X}_1^k;\hat{Y}_1^k)$ is the logarithm of the Radon-Nikodym density of the distribution of
$(\hat{X}_1^k,\hat{Y}_1^k)$ relative to the product of its marginal distributions.

Since $\hat{X}_1$ has the same distribution as $q(X_1^n)$, (\ref{eq.qproperty}) implies that
$E[l(\hat{X}_1)]\leq nP_{ave}-\delta$.   Thus, by the law of large numbers,
$\lim_{k\rightarrow\infty} P[\frac{1}{k} \sum_{j=1}^k l(\hat{X}_j)\geq nP_{ave} ]=0.$

Combining the facts from the preceding two paragraphs yields that, for $k$ sufficiently large,
$P[ (\hat{X}_1^k,\hat{Y}_1^k) \in G_k] \geq 1-\epsilon/2$, where $G_k$ is the following
subset of $A^k\times B^k$:
\begin{equation}
G_k  =  \{ \frac{1}{k} i_k(\hat{x}_1^k;\hat{y}_1^k) \geq n(\Cinfo - 3\epsilon) \}\cap  \{\frac 1 k \sum_{j=1}^k l(\hat{x}_j)\leq nP_{ave} \}.
\end{equation}
Therefore, by a variation of Feinstein's lemma, modified to take into account the average
power constraint (see below) for sufficiently large $k$ there exists a $(k,M,\epsilon)$ code
${\cal C}_{A,k}$ for the channel $\tilde{\nu}$ such that 
 $\log M \geq  kn(\Cinfo-4\epsilon)$ and 
each codeword satisfies the constraint
$ \frac{1}{k} \sum_{j=1}^k l(\hat{x}_j)\leq nP_{ave}$.

For any message value $j$ with $1\leq j \leq M$, passing the $j$th codeword
$\underline{a}$ of ${\cal C}_{A,k}$ through the channel with transition
kernel $\gamma$ generates a random codeword in $\ID_n(P_{peak})^k$, 
which we can also view as a random codeword $\underline{x}$ in $\IC^{nk}$. 
Since $|| \underline{x} ||_2^2 \leq \sum_{j=1}^k l(a_j) \leq nk P_{ave}$, the random codeword $\underline{x}$ satisfies the peak power constraint for the original channel, with probability one. 
Also, the average error probability for the random codeword is equal to the error probability for the codeword $\underline{a}$, which is at most $\epsilon$. 
Since the best case error probability is no larger than the average, there exists a deterministic choice of codeword $\underline{x}$, also satisfying the average power constraint and having error probability less than or equal to $\epsilon$.
Making such a selection for each codeword in  ${\cal C}_{A,k}$  yields an
$(nk, M, nkP_{ave}, P_{peak}, \epsilon)$ code for the original fading channel with
$\log(M) \geq nk(C_{info}-4\epsilon$).    Since $\epsilon > 0$ is arbitrary,
$\Cop(P_{ave}, P_{peak}) \geq \Cinfo(P_{ave}, P_{peak})$, as was to be proved.

It remains to give the modification of Feinstein's lemma used in the proof.  The lemma is stated
now using the notation of \cite[\S 12.2]{Gray}.   The lemma can be used above by
taking $A$, $A_o$, and $a$ in the lemma equal to $A^k$,
$A^k\cup\{\frac{1}{k}\sum_{j=1}^k l(\hat{x_j}) \leq nP_{ave}\}$,
and $nk(\Cinfo-3\epsilon)$, respectively.   The code to be produced is to have symbols from
a measurable subset $A_o$ of  $A$.  
\begin{lemma}[Modified Feinstein's lemma] \label{lemma.modFeinstein}
Given an integer $M$ and $a>0$ there exist
$x_j \in A_o; j=1,\ldots , M$ and a measurable partition ${\cal F}=\{\Gamma_j; j=1, \ldots, M\}$
of $B$ such that $\nu(\Gamma_j^c|x_j) \leq Me^{-a}+P_{XY}(\{i\leq a\}\cup (A_o^c\times B)).$
\end{lemma}
The proof of Lemma \ref{lemma.modFeinstein} is
the same as the proof given in \cite{Gray} with the set $G$ in \cite{Gray} replaced by the
set
$G=\{(x,y): i(x,y) \geq a,~ x\in A_o\}$ and with $\epsilon=Me^{-a}+P_{XY}(G^c)$.
The proof of Proposition  \ref{prop.CopCinfo} is complete.

\subsection{Proof of Proposition \ref{prop.peak_equiv}} \label{sec.peak_equivProof}
\begin{proof} 
For brevity, let $\alpha = \sup_{P_{ave} >  0}     \frac{\Cop(P_{ave},P_{peak})}{P_{ave}}$.
 We wish to prove that $C_p(P_{peak})=\alpha$.
The proof that $C_p(P_{peak}) \geq \alpha$ is identical to the analogous proof of \cite[Theorem 2]{Verdu90}.

To prove the converse, let $\epsilon >0$.  By the definition of  $C_p(P_{peak})$, for any
$\nu$ sufficiently large there exists an
$(n,M,\nu,P_{peak},\epsilon)$ code such that $\log M \geq \nu(C_p(P_{peak})-\epsilon)$.
Let $X_1^n$ be a random vector that is uniformly distributed over the set of $M$ codewords.
By Fano's inequality,  $I(X_1^n;Y_1^n) \geq (1-\epsilon)\log M - \log 2$. 
Setting $P_{ave}=\frac{\nu}{n}$, 
\begin{eqnarray}
\frac{1}{P_{ave}} \Cinfo(P_{ave} ,P_{peak}) &\geq & \frac 1 \nu I(X_1^n;Y_1^n) \\
		&\geq & (1-\epsilon)(C_p(P_{peak})-\epsilon) - \frac{\log 2}{\nu}.
\end{eqnarray}
Using the assumption that $\Cinfo(P_{ave},P_{peak})=\Cop(P_{ave},P_{peak})$ yields $\alpha \geq (1-\epsilon)(C_p(P_{peak})-\epsilon)-\frac{\log 2}{\nu}$.   
Since  $\epsilon$ can be taken arbitrarily small and $\nu$ can be taken arbitrarily large, $\alpha \geq C_p(P_{peak})$. This proves (\ref{eq:CapEnergyMutualInfoConnectVerdu0}).
Noting that $C_{op}(P_{ave},P_{peak})=C_{info}(P_{ave},P_{peak})$ by assumption, and using Definition \ref{defn:InfoTheoryDefnCapacity}, 
it is clear that (\ref{eq:CapEnergyMutualInfoConnectVerdu}) follows from (\ref{eq:CapEnergyMutualInfoConnectVerdu0}).  

Consider a time-varying fading channel modeled in discrete time as given in (\ref{eq:ChannelModel}). 
It is useful to consider for theoretical purposes a channel that is available for independent blocks of duration $n$. 
The fading process is time-varying within each block. 
However, the fading across distinct blocks is independent. 
Specifically, let $\widetilde{H}$ denote a fading process such that the blocks of length $n$, $(\widetilde{H}(1+kn), \ldots, \widetilde{H}(n+kn))$, indexed in $k\in \IZ$, are independent, with each having the same probability distribution as $\left(H(1),H(2),\ldots,H(n)\right)$. 
Let $C_{p,n}(P)$ denote  the capacity per unit energy of the channel  with fading process $\widetilde{H}$. 

From (\ref{eq:CapEnergyMutualInfoConnectVerdu0}), 
\begin{eqnarray}
C_p(P) &=& \sup_{n\geq 1} C_{p,n}(P) \label{eq:CapacityRelatedToDivergenceExpression}
\end{eqnarray}
The proof of Proposition \ref{prop.peak_equiv}  is completed as follows.  By its definition, $C_{p,n}(P)$ is
clearly monotone nondecreasing in $n$,  so that
$\sup_n C_{p,n} = \lim_{n\rightarrow\infty} C_{p,n}(P).$  Thus, the time-varying flat fading
channel is reduced to a block fading channel with independently fading blocks.  The
theory of memoryless channels in \cite{Verdu90} can be applied to the block fading channel,
yielding, for $n$ fixed, 
\begin{equation}
C_{p,n}(P) = \sup_{X \in {\ID}_n(P)} \frac{D(p_{Y|X}\|p_{Y|0})}{\|X\|_2^2} \label{eq:CapacityDivergenceExpressionFixedT}
\end{equation}
Equation (\ref{eq:CapacityDivergenceExpression}) follows from (\ref{eq:CapacityRelatedToDivergenceExpression}) and (\ref{eq:CapacityDivergenceExpressionFixedT}), and the proof is complete.
\end{proof}

\section{PROOF OF PROPOSITION \ref{prop.main} } \label{sec.proof}
The proof of Proposition \ref{prop.main} is organized as follows. 
The capacity per unit energy is expressed in (\ref{eq:CapacityDivergenceExpression}) as the supremum of a scaled divergence expression. To evaluate the  supremum, it is enough to consider codes with only one vector $X$ in the input alphabet, in addition to the all zero input vector.
In Section \ref{subsec.ReductionToOnOffSignaling}, ON-OFF signaling is introduced. It is shown that the supremum is unchanged if $X$ is restricted to be an ON-OFF signal; i.e. $X_i \in \{0,\sqrt{P} \}$ for each $i$.
In Section \ref{subsec.OptimalityOfTemporalOnOffSignaling},  the optimal choice of input vector $X$  is further characterized, and temporal ON-OFF signaling is introduced.
In Section \ref{subsec.IdentifyingTheLimit}, a well-known identity for the prediction error of a stationary Gaussian process is reviewed and applied to conclude the proof of Proposition \ref{prop.main}. 

\subsection{Reduction to ON-OFF Signaling} \label{subsec.ReductionToOnOffSignaling}
It is shown in this section that the supremum in (\ref{eq:CapacityDivergenceExpression}) is unchanged if $X$ is restricted to satisfy $X_i \in \{0,\sqrt{P} \}$ for each $i$. Equivalently, in every timeslot, the input symbol is either $0$ (OFF) or $\sqrt{P}$ (ON). We refer to this as ON-OFF signaling.

The conditional probability density \cite{MarzettaHochwald99} of the output  $n \times 1$ vector $Y$, given the input  vector $X$, is 
\begin{equation}
 p_{Y|X}(Y) = \frac{ \exp{\left( -tr{(I_n+\bar{X} \Sigma {\bar{X}}^{\dag})^{-1} YY^{\dag}} \right)} }{\pi^n .\det{(I_n+ \bar{X} \Sigma {\bar{X}}^{\dag})}} \label{eq:Cond_pdf}
\end{equation}
where $\bar{X}$ denotes the $n\times n$ diagonal matrix with diagonal entries given by $X$, and  $\Sigma$ is the covariance matrix of the random vector $(H(1), \ldots, H(n))^T$. The divergence expression is simplified by integrating out the ratio of the probability density functions.
\begin{eqnarray}
D(p_{Y|X}\|p_{Y|0}) &=& \int  p_{Y|X}(Y) \log\left( \frac{ p_{Y|X}(Y) }{ p_{Y|0}(Y) } \right)  \, dY \nonumber \\
&=&\int p_{Y|X}(y)\left\{ -tr (I_n + \bar{X}\Sigma \bar{X}^{\dag})^{-1} YY^{\dag} \right. \nonumber\\
		& & \left. + tr(YY^{\dag})-\log{ \det{(I_n + \bar{X}\Sigma \bar{X}^{\dag})}} \right\} \, dY \nonumber \\
&=& - tr(I_n + \bar{X}\Sigma \bar{X}^{\dag})^{-1}(I_n + \bar{X}\Sigma \bar{X}^{\dag}) + tr(I_n + \bar{X}\Sigma \bar{X}^{\dag}) \nonumber\\
 & & - \log{ \det{(I_n + \bar{X}\Sigma \bar{X}^{\dag})}} \nonumber \\
&=& tr(\bar{X}\Sigma \bar{X}^{\dag}) - \log{ \det{ (I_n + \bar{X}\Sigma \bar{X}^{\dag})}} \nonumber
\end{eqnarray}
Since the correlation matrix $\Sigma$ of the fading process is normalized, it has all ones on the main diagonal. Thus $tr(\bar{X}\Sigma \bar{X}^{\dag}) = \sum_{i=1}^{n} |X_i|^2$, so  
\begin{eqnarray}
C_p(P) &=& \lim_{n\rightarrow \infty} \sup_{X \in {\ID}_n(P)} \frac{\sum_{i=1}^n |X_{i}|^2-\log \det (I+\bar{X} \Sigma {\bar{X}}^{\dag})}{\|X\|_2^2}\label{eq:sup_C}\nonumber \\
&=& 1-\lim_{n\rightarrow \infty} \inf_{X \in {\ID}_n(P)} \frac{\log\det(I+\bar{X} \Sigma {\bar{X}}^{\dag})}{\|X\|_2^2}. \label{eq:C1}
\end{eqnarray}

Here $(X_1,\ldots,X_n)$ takes values over deterministic complex $n\times 1$ vectors with $|X_i|^2\leq P$. 
Consider $\bar{X} = R\exp(j\Theta)$, where $R$ is a nonnegative diagonal matrix, and $\Theta$ is diagonal with elements {$\Theta_i \in [0, 2\pi]$}. Using $\det(I+AB) = \det(I+BA)$ for any $A,B$, we get
\begin{eqnarray*}
\det(I + \bar{X}\Sigma {\bar{X}}^{\dag})&=& \det(I+ {\bar{X}}^{\dag} \bar{X} \Sigma)\\
&=& \det(I+R^2\Sigma)
\end{eqnarray*}
Hence we can restrict the search for the optimal choice of the matrix $\bar{X}$ (and hence of the input vector signal $X$) to real nonnegative  vectors. So, $\log\det(I_n + \bar{X}\Sigma {\bar{X}}^{\dag})=$ $\log\det(I_n+{\bar{X}}^2 \Sigma)$.

Fix an index $i$ with $1\leq i \leq n$. Note that $\det(I_n + X^2 \Sigma)$ is linear in $X_i^2$. Setting $X_i^2 = u$, the expression to be minimized in (\ref{eq:C1}) can be written as a function of $u$ as
\begin{equation}
f(u) =  \frac{\log \det(I_n+X^2 \Sigma)}{\| X\|_2^2} = \frac{\log(a+bu)}{c+u} , ~0\leq u \leq P \label{eq:PositiveFunction}
\end{equation}
for some non-negative $a$, $b$ and $c$. 
Since $\Sigma$ is positive semidefinite, all the eigenvalues of $I_n+X^2 \Sigma$ are greater than or equal to $1$. Thus both the numerator and the denominator of (\ref{eq:PositiveFunction}) are nonnegative. The second derivative of $f(u)$ is given by 
\begin{eqnarray*}
f''(u)  =  -\frac{2}{c+u} \cdot f'(u)-\frac{b^2}{(c+u)(a+bu)^2}.
\end{eqnarray*}
So, $f(u)$ has no minima and at the most one maximum in the interval $[0,P]$. Since $u$ is constrained to be chosen from the set $[0,P]$,  $f(u)$ (and hence the function of interest) reaches its minimum value only when $u$ is either $0$ or $P$. This narrows down the search for the optimal value of $X_i^2$ from the interval $[0,P]$ to the set $\{0,P\}$ for all $i\in \{1,\ldots,n\}$. Restricting  our attention to values of $X$ with $X_i \in \{0,P\}$, we get the following expression for capacity per unit energy:
\begin{equation}
C_p(P) = 1 - \inf_n \inf_{ \begin{array}{c}
	\{0,\sqrt{P} \} \mbox{ valued signals}\\
	\mbox{with support in } \{1,\ldots, n\} \end{array} }\frac{ \log \det( I_n + X^{\dag}X\Sigma)}{\|X\|_2^2} \label{eq:CapacityWithBlockLength}
\end{equation}
Consider the expression \[ \frac{ \log \det( I_n + X^{\dag}X\Sigma)}{\|X\|_2^2}
\] Here, $n$ is the block length, while $X$ is the input signal vector, with $X_i \in \{0,\sqrt{P}\}$.  Having a certain block length and an input signal vector has the same effect on the above expression as having a greater block length and extending the input signal vector by appending the required number of zeros.
 So, the  expression does not depend on the block length $n$, as long as $n$ is large enough to support the input signal vector $X$.

Since the block length $n$ does not play an active role in the search for the optimal input signal, (\ref{eq:CapacityWithBlockLength}) becomes
\begin{equation}
C_p(P) = 1 - \inf_{k\geq 1} \inf_{  \begin{array}{c}
 	\{0,\sqrt{P} \} \mbox{ valued signals}\\
	\mbox{with energy } kP \end{array}} \frac{ \log \det( I + X^{\dag}X\Sigma)}{kP}. \label{eq:C2}
\end{equation}

From here onwards, it is implicitly assumed that, for any choice of input signal $X$, the corresponding block length $n$ is chosen large enough to accommodate $X$. 

\subsection{Optimality of Temporal ON-OFF Signaling} \label{subsec.OptimalityOfTemporalOnOffSignaling}
We use the conventional set notation of denoting the intersection of sets $A\cap B$ by $AB$ and $A$'s complement by $A^c$.

Consider the random process $Z$
\begin{equation}
Z_k = \sqrt P H_k + W_k ~\forall~k \label{eq:newmodel}
\end{equation}
In any timeslot $k$, if the input signal for the channel (\ref{eq:ChannelModel}) is $\sqrt{P}$, then the corresponding output signal is given by (\ref{eq:newmodel}). Otherwise, the output signal is just the white Gaussian noise term $W_k$. A $\{0,\sqrt{P}\}$-valued signal $X$ with finite energy can be expressed as $X = \sqrt{P} I_A$, where $A$ is the support set of $X$ defined by $A = \{i:X_i \neq 0\}$, and $I_A$ denotes the indicator function of $A$. Thus, $A$ is the set of ON times of signal $X$, and $|A|$ is the number of ON times of $X$. 
\begin{defn}
Given a finite subset $A$ of ${\IZ}_+$, let
\begin{equation*}
\alpha(A)  = \left\{ \begin{array}{cc}
 			\log \left[\det(I+P diag(I_A) \Sigma)\right] &  \mbox{ if } A \neq \emptyset\\
 			0		 & \mbox{ if } A = \emptyset \end{array} \right.
\end{equation*}
Further, for any two finite sets $A$,$B$  $\subset \IZ$, define $\alpha(A|B)$ by \[ \alpha(A|B) = \alpha(A\cup B) - \alpha(B) \]
\end{defn}
It is easy to see that for $A \neq \emptyset$, \[ \alpha(A) = h(Z_i: i\in A) - |A| \log(\pi e)\] where $h(.)$ is the differential entropy of the specified random variables. Note that the term $- |A| \log(\pi e)$ in the definition of $\alpha$ is linear in $|A|$.  Also, $\alpha(.|.)$ is related to the conditional differential entropies of the random variables corresponding to the sets involved. Specifically, \[ \alpha(A|B) = h(Z_i: i\in AB^c | Z_j : j \in B) - |AB^c| \log(\pi e)\]
We are interested in characterizing the optimal signaling scheme that would achieve the infima in (\ref{eq:C2}). Since the input signal is either $\sqrt{P}$ or $0$, the expression inside the infima in  (\ref{eq:C2}) can be simplified to $\frac{\alpha(A)}{P \cdot |A|}$, where $A$ is the set of indices of timeslots where the input signal is nonzero. 
Thus, the expression for capacity per unit energy reduces from (\ref{eq:C2}) to
\begin{equation}
C_p(P)	= 1-  \inf_{ 0< |A| <\infty } \frac{\alpha(A)}{P \cdot |A|} . \label{eq:C3}
\end{equation}

\begin{lemma}
The functional $\alpha$ has the following properties.
 \begin{enumerate}
 \item $\alpha(\emptyset) = 0$ 
 \item If $C \subset D$, then $\alpha(C)\leq \alpha(D)$. Consequently, $\alpha(A) \geq 0 ~\mbox{for each} ~ A \neq \emptyset $.
 \item Two alternating capacity property: $\alpha(A \cup B) + \alpha(AB) \leq \alpha(A) +\alpha(B)$
 \item $\alpha(B) = \alpha(B+k) ~\mbox{for each} ~k \in \IZ$
 \end{enumerate} \label{lem.4properties}
\end{lemma}
\begin{proof} 
The first property follows from the definition. 
From the definition of $\alpha(.|.)$, $\alpha(D) = \alpha(DC^c | C)+\alpha(C)$, so the second property is proved if $\alpha(DC^c | C) \geq 0$. But, 
\begin{eqnarray*}
\alpha(DC^c | C) &=&  h(Z_i: i\in DC^c | Z_j : j \in C) - |DC^c| \log(\pi e) \\
&\stackrel{(a)}{\geq}& h(Z_i: i\in DC^c | Z_j: j \in C,~H_j: j \in DC^c) - |DC^c| \log(\pi e) \\
&\stackrel{(b)}{=}& h(W_{ \{DC^c \} })- |DC^c|\cdot \log(\pi e) = 0
\end{eqnarray*}
where $W_{\{DC^c\} }$ denotes the vector composed of the random variables $\{W_i : ~i~ \in~ DC^c\}$. Here, $(a)$ follows from the fact that conditioning reduces differential entropy, while $(b)$ follows from the whiteness of the Gaussian noise process $W$.

Since the term $-|A| \log(\pi e)$ in the definition of $\alpha$ is linear in $|A|$, the third property for $\alpha$ is equivalent to the same property for the set function $A \to h(Z_i: i~\in~A)$. This equivalent form of the third property is given by \[ h(Z_i:i \in AB^c | Z_j: j \in B) \leq h(Z_i:i \in AB^c | Z_j:j \in AB )
\] But this is the well known property that conditioning on less information increases differential entropy. This proves the third property. The fourth part of the proposition follows from the stationarity of the  random process $H$.
\end{proof}

The only properties of $\alpha$ that are used in what follows are the properties listed in the above lemma; i.e., in what follows, $\alpha$ could well be substituted with another functional $\beta$, as long as $\beta$ satisfies the properties in Lemma \ref{lem.4properties}.
\begin{lemma}
Let $A,B \subset \IZ $ be finite disjoint nonempty sets. Then
\begin{equation}
\frac{\alpha(A\cup B)}{|A\cup B |} \leq \frac{\alpha(A)}{| A|} ~~\Leftrightarrow ~~ \frac{\alpha(B|A)}{|B|} \leq \frac{\alpha(A\cup B)}{|A\cup B|} ~~\Leftrightarrow ~~ \frac{\alpha(B|A)}{|B|} \leq \frac{\alpha(A)}{|A|} 
\end{equation} \label{lem.EquivalentAlpha}
\end{lemma} 
\begin{proof} Trivially,
\[ \frac{\alpha(A)+\alpha(B|A)}{|A|+|B|} = \frac{\alpha(A\cup B)}{|A\cup B|} \]
Each individual term of the numerators and denominators in the above equation is nonnegative. 
Note that for $a,b >0$ and $c,d \geq 0$:
\[ \frac{c+d}{a+b} \leq \frac{c}{a} \Leftrightarrow \frac{d}{b}\leq \frac{c+d}{a+b} \Leftrightarrow \frac{d}{b}\leq \frac{c}{a}\]
Letting $a = |A|, b = |B|, c = \alpha(A), d = \alpha(B|A)$, the lemma follows.
\end{proof}

Let $E,F$ be two nonempty subsets of $\IZ$ with finite cardinality.  
$E$ is defined to be \emph{better} than $F$ if \[ \frac{\alpha(E)}{|E|} \leq \frac{\alpha(F)}{|F|}
\]
\begin{lemma}
Let $A ~\subset ~\IZ$ be  nonempty with  finite cardinality. Suppose  \( \frac{\alpha(A)}{|A|} \leq \frac{ \alpha(C)}{|C|} \) for all  nonempty proper subsets $C$ of $A$. Suppose $B$ is a set of finite cardinality such that \( \frac{\alpha(B)}{|B|} \leq \frac{\alpha(A)}{|A|} \). So, $B$ is better than $A$, and $A$ is better than any nonempty subset of $A$. Then, for any integer $k$,
\[ \frac{\alpha(\widetilde{A} \cup B )}{|\widetilde{A} \cup B |} \leq \frac{ \alpha(A)}{|A|} \]
where $\widetilde{A}=A+k$, i.e., $\widetilde{A}$ is obtained by incrementing every element of $A$ by $k$.\label{lem.PlusOne}
\end{lemma}
\begin{proof}
It suffices to prove the result for $\widetilde{A} = A$, for otherwise $B$ can be suitably translated. Let $ D = B A^c$ and $\widetilde{D} = B A$. The set $B$ is better than $A$, and hence better than any subset of $A$. In particular, $B$ is better than $\widetilde{D}$. $B$ is the union of the two disjoint sets $D$ and $\widetilde{D}$. Applying Lemma \ref{lem.EquivalentAlpha}  to  $\widetilde{D}$ and $D$ yields $\frac{\alpha(D|\widetilde{D})}{|D|}  \leq \frac{\alpha(B)}{|B|}$. 
Since, $\frac{\alpha(B)}{|B|} \leq \frac{\alpha(A)}{|A|}$, it follows that $\frac{\alpha(D|\widetilde{D})}{|D|} \leq \frac{\alpha(A)}{|A|}$. 
The fact that $\widetilde{D}$ is a subset of $A$, and the second property of $\alpha$ applied to $A$ and $D$, together imply that $\frac{\alpha(D|A)}{|D|} \leq \frac{\alpha(D|\widetilde{D})}{|D|} \leq \frac{\alpha(A)}{|A|}$. 
Consequently, application of Lemma \ref{lem.EquivalentAlpha} to the disjoint sets $A$ and $D$ yields that $\frac{\alpha(A \cup D )}{|A \cup D |} \leq  \frac{ \alpha(A)}{|A|}$ which is equivalent to the desired conclusion.
\end{proof}

\begin{prop} The following holds.\[
\inf_{A:A ~finite} \frac{\alpha(A)}{|A|} = \lim_{n\to \infty} \frac{\alpha(\{1,\ldots,n\})}{n} \] \label{prop:M}
\end{prop} 
\begin{proof}
Let $\epsilon >0$. Then, there exists a finite nonempty set $A_1^*$ with \[ \frac{\alpha(A_1^*)}{|A_1^*|} \leq \alpha^* +\epsilon \mbox{ where }\alpha^* = \inf_{A:A ~finite} \frac{\alpha(A)}{|A|}\]
Let $A^*$ be a smallest cardinality nonempty subset of $A_1^*$ satisfying the inequality \[ \frac{\alpha(A^*)}{|A^*|} \leq \alpha^* +\epsilon \]
Then \[ \frac{\alpha(A^*)}{|A^*|} \leq \frac{\alpha(A)}{|A|} \mbox{ for any $A\subset A^*$ with } A \neq \emptyset\]
Let $S_1 = A^*$. For $k > 1$, let $S_k = A^* \cup (A^*+1) \cup \ldots \cup (A^*+k-1)$. For $k\geq 1$, let $T_k $ be the claim that $S_k$ is better than $A^*$.
The claim $T_1$ is trivially true. For the sake of argument by induction, suppose $T_k$ is true for some $k\geq 1$. 
Choose $B = S_k$, and $\widetilde{A} = A^* + k$ and apply Lemma \ref{lem.PlusOne}. This proves the claim $T_{k+1}$. Hence, by induction, $T_k$ is true $\forall ~ k \in N$.
So, for any $k \in N$, $A^* \cup (A^*+1) \cup \ldots \cup (A^*+k-1)$ is better than $A^*$.
Roughly speaking, any gaps in the set $S_k$ are removed with $k$. 
So, for every $\epsilon$, we can find an $n_{\epsilon}$ so that for all $n\geq n_{\epsilon}$, $A_{n} =\{1,\ldots,n\}$ satisfies: 
\[ \alpha^* \leq \frac{\alpha(A_n)}{|A_n|} \leq \alpha^* + \epsilon  ~\forall n\geq n_{\epsilon}
\] 
Hence the proposition is proved.
\end{proof}

Equation (\ref{eq:C3}) and Proposition \ref{prop:M} imply that the capacity per unit energy is given by the following limit:
\begin{eqnarray}
C_p(P)	& = & 1 - \lim_{n\to \infty} \frac{\alpha(\{1,\ldots,n\})}{nP}\nonumber \\
 & = & 1 - \lim_{n\to \infty} \frac{\log \det(I_n + P\Sigma_{n\times n} )}{nP} \label{eq:C4}
\end{eqnarray}
		At this point, it may be worthwhile to comment on the structure of a signaling scheme for achieving $C_p(P)$ for the original channel. 
The structure of codes achieving capacity per unit energy for a memoryless channel with a zero cost symbol is outlined in \cite{Verdu90}. 
This, together with Propositions \ref{prop.peak_equiv} and \ref{prop:M}, show that 
$C_p(P)$ can be asymptotically achieved by codes where each codeword $W$ has the following structure for constants $N,T,d$ with $N \gg 1$ and $1\ll T\ll d$:
\begin{itemize}
		\item Codeword length is $N(T+d)$.
		\item $W_i\in \{0,\sqrt{P}\}$ for all $0\leq i < N(T+d)$.
		\item $W_i$ is constant over intervals of the form $[k(T+d), k(T+d)+T-1]$.
		\item $W_i$ is zero over intervals of the form $[k(T+d)+T, (k+1)(T+d)-1]$.
		\item $\sum_i W_i^2 \ll NT$
\end{itemize}
So, the vast majority of codeword symbols are OFF, with infrequent long bursts of ON symbols. This is referred to as temporal ON-OFF signaling.

\subsection{Identifying the limit} \label{subsec.IdentifyingTheLimit}
We shall  show that (\ref{eq:C4}) is equivalent to (\ref{eq:EnergyBound})-(\ref{eq:mainprop_I}).

Let $Z$ be the process defined by (\ref{eq:newmodel}). Consider the problem of estimating the value of $Z(0)$ by observing the previous $n$  random variables $\{Z(k): -n\leq k<0\}$ such that the mean square error is minimized. Since $Z$ is a proper complex Gaussian process, the minimum mean square error estimate  of $Z(0)$ is linear \cite[Chapter IV.8 Theorem 2]{GihmanSkorohod}, and it is  denoted by $\widetilde{Z}(0|-1,\hdots,-n)$. Let $\sigma^2_{0|-1,\hdots,-n} = E\left[\left|Z(0)~ - ~ \widetilde{Z}(0|-1,\hdots,-n)\right|^2\right]$ be the mean square error. Let $D_n$ denote the determinant $\det( I_{n} + P \Sigma_{n} )$.
Note that $D_n \geq 1$ for all $n$ since $\Sigma_n$, being an autocorrelation matrix, is positive semidefinite for all $n$.
\begin{lemma} \label{lem:NStepLogDetLemma}
The minimum mean square error in predicting $Z(0)$ from $\{Z(k): -n\leq k<0\}$ is given by
\begin{equation*}
\sigma^2_{0|-1,\hdots,-n} = \frac{D_{n+1}}{D_n} \nonumber 
\end{equation*}
\end{lemma}
\begin{proof}
The random variables $\underline{Z}_{-n}^0 = \{ Z(k): -n\leq k \leq 0\}$ are jointly proper complex Gaussian and have the following  expression for differential entropy. \[ h(\underline{Z}{}_{-n}^0) = \log((\pi e)^{n+1}D_{n+1} )
\]
The differential entropy of $\widetilde{Z}(0|-1,\hdots,-n)$ is the conditional entropy  $h\left(Z(0)|\underline{Z}_{-n}^{-1}\right)$, which can be expressed in terms of $D_{n+1}$ and $D_n$ as follows.
\begin{eqnarray}
h\left(\widetilde{Z}(0|-1,\hdots,-n)\right) & = & h\left(Z(0)| \underline{Z}_{-n}^{-1} \right) \nonumber \\
			   	& \stackrel{(a)}{=} & h(\underline{Z}_{-n}^0) - h(\underline{Z}_{-n}^{-1}) \nonumber\\
				& = & \log((\pi e)^{n+1} D_{n+1}) - \log((\pi e)^n D_n) = \log \left( \pi e \frac{D_{n+1}}{D_n} \right) \nonumber 
\end{eqnarray}
where $(a)$ follows from the fact that, for any two random vectors $U$ and $V$, the conditional entropy $h(U|V) = h(U,V) - h(V)$.
Since $\widetilde{Z}(0|-1,\hdots,-n)$ is a linear combination of proper complex jointly Gaussian random variables, it is also proper complex Gaussian. Hence, its differential entropy is  given by 
\begin{equation}
h(\widetilde{Z}(0|-1,\hdots,-n)) = \log(\pi e ~ \sigma^2_{0|-1, \ldots, -n}) \label{eq:EntropyRateExpressionFrom1Ton}
\end{equation}
The lemma follows by equating the above two expressions for the differential entropy of \\
$\widetilde{Z}(0|{-1},\hdots,-n)$.

\end{proof}

 The $n$-step mean square prediction error $\sigma^2_{0|-1,\hdots,-n}$ is non-increasing in $n$, since projecting onto a larger space can only reduce the mean square error. 
So, the prediction error of $Z(0)$ given $(Z(-1), Z(-2), \hdots)$ is the limit of the sequence of the $n$-step prediction errors.
\begin{equation}
\lim_{n\to \infty} \sigma^2_{0|-1,\hdots,-n} = \sigma^2_{0|-1,-2\hdots} \label{eq:PredictionErrorAsLimitOfSequence}
\end{equation}
It follows from Lemma \ref{lem:NStepLogDetLemma} that the ratio of the determinants, $D_{n+1}/D_n$  converges to the prediction error of $Z(0)$ given $(Z(-1), Z(-2), \hdots)$.
\begin{equation*}
\lim_{n\to \infty} \frac{D_{n+1}}{D_n} = \sigma^2_{0|-1,-2\hdots} \nonumber 
\end{equation*}
The sequence $D_n^{1/n}$ also converges, since $\frac{D_{n+1}}{D_n}$ converges, and it converges to the same limit as $D_{n+1}/D_{n}$.
\begin{equation}
\lim_{n\to \infty} D_n^{1/n} = \sigma^2_{0|-1,-2\hdots} \label{eq:DetNthRoot}
\end{equation}

Let $(F_Z(\omega): -\pi \leq \omega < \pi)$ be the spectral distribution function of the process $Z$. Returning to the prediction problem, the mean square prediction error $\sigma^2_{0|-1,-2\hdots}$ can be expressed in terms of the density function of the absolutely continuous component of the power spectral measure of the process $Z$ \cite[Chapter XII.4 Theorem 4.3]{Doob53}.
\begin{equation}
\sigma^2_{0|-1,-2\hdots} = \exp\left( \int_{-\pi}^{\pi} \log F_Z^{'}(\omega) \frac{d\omega}{2\pi} \right) \label{eq:MeanSquareErrorInTermsOfF_Z}
\end{equation}

From (\ref{eq:DetNthRoot}), we know that the $\log \det$ term in the capacity per unit energy expression (\ref{eq:C4}) converges to the $\log (\sigma^2_{0|-1,-2\hdots})$. 
Equation (\ref{eq:MeanSquareErrorInTermsOfF_Z}) relates the mean square prediction error of a  wide sense stationary process to the spectral measure of the process. 
This lets us simplify the $\log \det$ term into an integral involving the power spectral density $S(\omega)$ of the fading process $H$.
We state and prove the following lemma.
\begin{lemma} \label{lem:LogDetLemma}
\begin{equation*}
I(P) = \lim_{n \to \infty} \frac{1}{n} \log \det(I_n + P\cdot \Sigma_n)
\end{equation*}
\end{lemma}
\begin{proof}
Let $( F_H(\omega), F_W(\omega): -\pi \leq \omega < \pi)$ be the spectral distribution functions of the processes $H$, and $W$ respectively. \[ F_Z(\omega) = P \cdot F_H(\omega) + F_W(\omega)
\] 

The density $F_H^{'}$ of the absolutely continuous part of the power spectral measure of the fading process $H$ is given by $S(\omega)$. Since $W$ is white Gaussian, its spectral distribution $F_W$ is absolutely continuous with density $1$. 
Hence the density $F_Z^{'}$ of the absolutely continuous component  of $F_Z$ is given by 
\begin{equation}
F_Z^{'}(\omega) = 1+P\cdot S(\omega) \label{eq:PSD}
\end{equation}

The expression for the mean square prediction error $\sigma^2_{0|-1,-2\hdots}$ in (\ref{eq:MeanSquareErrorInTermsOfF_Z}) involves the density function $F_Z^{'}$.
Substituting the density function by the expression in (\ref{eq:PSD}), we get
\begin{equation}
\sigma^2_{0|-1,-2\hdots} = \exp \left( \int_{-\pi}^{\pi} \log(1+P\cdot S(\omega)) \frac{d\omega}{2\pi}\right) \label{eq:MMSEIntegral}
\end{equation}
From (\ref{eq:mainprop_I}) and (\ref{eq:MMSEIntegral}), it follows that 
\begin{equation}
\sigma^2_{0|-1,-2\hdots} = e^{I(P)} \label{eq:MMSEExponentOfI}
\end{equation}
The lemma follows from equating the expressions for $\sigma^2_{0|-1,-2\hdots}$ in  (\ref{eq:DetNthRoot}) and (\ref{eq:MMSEExponentOfI}).
\end{proof}

Let $\widehat{I}(P)$ be the mutual information rate between the fading process $H$ and $Z$, when the input is identically $\sqrt{P}$, as modeled in (\ref{eq:newmodel}). It is interesting to note that  $\widehat{I}(P)$  is related to $I(P)$ in the following manner. 
\begin{eqnarray}
\widehat{I}(P) &=& \lim_{n \to \infty} \frac{1}{n} I(Z_{-1} \hdots Z_{-n}; H_{-1} \hdots H_{-n})\\
			&=& \lim_{n \to \infty} \frac{1}{n} h(Z_{-1} \hdots Z_{-n}) - \frac{1}{n} h(W_{-1} \hdots W_{-n})  \label{eq:MutualInformationRateAsASum}
\end{eqnarray}
The first term in (\ref{eq:MutualInformationRateAsASum}) is the entropy rate $h_Z$ of the process $Z$. Following (\ref{eq:EntropyRateExpressionFrom1Ton}), (\ref{eq:PredictionErrorAsLimitOfSequence}) and (\ref{eq:MMSEIntegral}), $h_Z$ is given by    
\begin{eqnarray}
h_Z = \log( \pi e) + I(P) \label{eq:EntropyRateOfZWhenON}
\end{eqnarray}
The second term in (\ref{eq:MutualInformationRateAsASum}) is the entropy rate of the white Gaussian process $W$, given by $\log(\pi e)$. 
From (\ref{eq:MutualInformationRateAsASum}) and (\ref{eq:EntropyRateOfZWhenON}), it follows that the mutual information rate $\widehat{I}(P)$ is equal to $I(P)$.

 We briefly outline an alternative way to prove Lemma \ref{lem:LogDetLemma} in Appendix \ref{app:LogDetLemmaAlternateProof}. 
Additional material on the limiting distribution of eigenvalues of Toeplitz matrices can be found in \cite[Section 8.5]{Gallager68}. 
Lemma \ref{lem:LogDetLemma} is used to simplify the capacity expression in (\ref{eq:C4}). Using the above simplification, the capacity per unit energy is given by
\begin{equation}
C_p(P)  =  1 - \frac{I(P)}{P} \label{eq:C5}
\end{equation}
This proves Proposition \ref{prop.main}.

\section{EXTENSION TO CHANNELS WITH SIDE INFORMATION: Proof of Proposition \ref{prop.CSImain}} \label{sec.extensionCSI}
Considering CSI at the receiver as part of the output, the channel output can be represented by 
\begin{equation}
\breve{Y}(k) = \left( \begin{array}{c} X(k)H(k) + W(k) \\
				H(k) \end{array} \right).
\end{equation}
Since $H$ and $W$ are stationary and weakly mixing, and 
the processes $H$, $W$ and $X$ are mutually independent, it can be shown that the above channel is stationary and ergodic. 
Propositions \ref{prop.CopCinfo} and \ref{prop.peak_equiv} can be extended to hold for the above channel. Recall that $C_p^{coh}(P)$ denotes the capacity per unit energy of this channel under a peak constraint $P$. 

Let $\widetilde{H}$ denote a fading process where blocks of length $T$, $( \widetilde{H}(1+kT), \ldots, \widetilde{H}(T+kT))$, indexed in $k \in Z$, are independent, and each block has the same distribution as $\left(H(1),H(2),\ldots,H(T)\right)$. 
A channel with the above fading process and with CSI at the receiver can be represented by  
\begin{equation}
	\hat{Y}(k) = \left[ \begin{array}{c}	
                 X(1+kT)\widetilde{H}(1+kT)+W(1+kT) \\
                 \vdots \\
                 X(T+kT)\widetilde{H}(T+kT)+W(T+kT) \\
                 \widetilde{H}(1+kT) \\
                 \vdots \\
                 \widetilde{H}(T+kT) \end{array} \right], \label{eq:BlockFadingCoherenceChannel}
\end{equation}
with  input $\left( X(1+kT), \hdots, X(T+kT) \right)$ and output $\hat{Y}(k)$.  
Let $C_{p,T}^{coh}(P)$ denote the capacity per unit energy of this channel. 
Using a simple extension of (\ref{eq:CapacityRelatedToDivergenceExpression}) in  Section \ref{sec.peak_equivProof},
it can be shown that 
\begin{equation}
C_p^{coh}(P)  = \lim_{T\rightarrow \infty} C_{p,T}^{coh}(P) \label{eq:EqualCapacitiesCoherent}
\end{equation}
\begin{lemma} \label{lem.CapacityPerUnitEnergyOfBlockChannelWithCSI}
For each $P > 0$ and $T > 0$, 
\begin{equation}
C_{p,T}^{coh}(P)  = 1  \label{eq:CapacityPerUnitEnergyOfBlockChannelWithCSI} 
\end{equation}	
\end{lemma}
For a proof of Lemma \ref{lem.CapacityPerUnitEnergyOfBlockChannelWithCSI}, see Appendix \ref{app.CapacityPerUnitEnergyOfBlockChannelWithCSIProof}.  
Proposition \ref{prop.CSImain} follows from  (\ref{eq:EqualCapacitiesCoherent}) and Lemma \ref{lem.CapacityPerUnitEnergyOfBlockChannelWithCSI}.

\section{EXTENSION TO CONTINUOUS TIME: Proof of Proposition 
\ref{prop.continuous}} \label{sec.ExtensionContinuousTime}
The proof of Proposition \ref{prop.continuous} is organized as follows. The capacity per unit energy with peak constraint of the given continuous-time channel is shown to be the limit of that of a discrete-time channel, suitably constructed from the original continuous-time channel. 
A similar approach is used in \cite{Wyner88} in the context of direct detection photon channels. 
The limit is then evaluated to complete the proof.

Recall that the observed signal (\ref{eq:ChannelModelContinuous}) is given by $$Y(t) = H(t)X(t)+W(t), ~0\leq t\leq T,$$ where $X(t)$ is the input signal. Here, $W(t)$ is a complex proper Gaussian white noise process. The fading process $H(t)$ is a stationary proper complex Gaussian process.
The observed integral process (\ref{eq:Vrepresentation}) is then $$V(t) = \int_0^t H(s)X(s)ds + \eta(t), ~0\leq t \leq T,$$ where $\eta$ is a standard proper complex Wiener process with autocorrelation function $E[\eta(s)\overline{\eta(t)}]=\min\{s,t\}$.

For an integer $J\geq 1$, a codeword $X$ is said to be in class $J$ if $X$ is
constant on intervals of the form $(i2^{-J}, (i+1)2^{-J}]$. 
A codeword is said to be a finite class codeword if it is in class $J$ for some finite $J$. 
Note that a class $J$ codeword is also a class $J'$ codeword for any $J'\geq J$. 
Given an integer $K\geq 1$ a decoder is said to be in class $K$ if it makes its decisions based only on the observations $Y^K=(Y^K(i): i\geq 0)$, where
\begin{equation}
Y^K(i) = \int_{i2^{-K}}^{(i+1)2^{-K}} Y(t) dt = V((i+1)2^{-K})-V(i2^{-K}).
\end{equation}
Note that a class $K$ coder is also a class $K'$ coder for any $K' \geq K$.
Let $C^{J,K}_p(P)$ denote the capacity per unit energy with peak constraint
$P$ when only class $J$ codewords and class $K$ decoders are permitted to be used.

Observe that, taking $J=K$, if a code consists of class $K$ codewords and if a class $K$ decoder is used, then the communication system is equivalent to a discrete time system.  
Therefore, it is possible to identify $C^{K,K}_p(P)$ using Proposition \ref{prop.main}.

Note that $C_p(P) \geq C_p^{J,K}(P)$ for any finite $J$ and $K$, because imposing restrictions on the codewords and decoder cannot increase capacity.  For the same reason, $C_p^{J,K}(P)$ is non-decreasing in $J$ and in $K$.  
Letting $J=K$ and taking the limit $K\rightarrow \infty$ yields
\begin{equation}  \label{eq.CPinequal}
C_p(P)  \geq \lim_{K\rightarrow\infty}  C^{K,K}_p(P).
\end{equation}
The proof is completed by showing that $C_p(P)  = \lim_{K\rightarrow\infty} C_p^{K,K}(P)$, and then identifying the limit on the right as the expression for capacity per unit energy given in the proposition. 

\begin{lemma} \label{lemma.cppsup}  $C_p(P)=\sup_{X\in L^2[0,\infty) } \frac{ D(P_{V|X}||P_{V|0})  }{||X||^2_2} $.
\end{lemma}
\begin{proof} The continuous-time channel is equivalent to a discrete-time abstract alphabet
channel with input alphabet $L^2[0,1]$ and output alphabet $C_0[0,1]$, the space of complex-valued
continuous functions on the interval $[0,1]$ with initial value zero.
For convenience, let $T$ be a positive integer. 
Then an input signal $(X(t):0\leq t \leq T)$ is equivalent to
the discrete-time signal $(X_0,\ldots, X_{T-1})$, where $X_i$ are functions on $L^2[0,1]$ defined by $X_i(s)=( X(s+i): 0\leq s \leq 1)$.
Similarly the output signal $(V(t): 0\leq t \leq T) $ is equivalent to the discrete-time signal
$(V_0,\ldots, V_{T-1})$, where $V_i(s)=(V(s+i)-V(i): 0\leq s \leq 1)$.  
Propositions  \ref{prop.CopCinfo} and  \ref{prop.peak_equiv}  generalize to this discrete-time channel with the same proofs,  yielding the lemma.
\end{proof}
\begin{lemma}  \label{lemma.Dcontinuity}
The  divergence $D(P_{V|X} || P_{V|0})$ as a function of $X$, which maps $L^2[0,\infty)$ to
$[0,\infty)$,  is lower semi-continuous.
\end{lemma}
\begin{proof}
Let $P_{V|X,H}$ denote the distribution of $V$ given $(X,H)$. 
Let $P_{V|\tilde{X},H}$ be defined  similarly. 
Given $(X,H)$, as shown in (\ref{eq:Vrepresentation}), $V$ is simply given by the integral
of a known signal $XH$ plus a standard proper complex Wiener process.
Consequently, the well-known Cameron-Martin formula for likelihood ratios can be
 used to find $D(~P_{V|X,H} ~|| ~P_{V|\tilde{X},H}~ )=  || XH-\tilde{X}H ||^2_2$. 
The measure $P_{V|X}$ is obtained from $P_{V|X,H}$ by integrating out $H$. 
Namely, for any Borel set $A$ in the space of $V$,  $P_{V|X}[A] = E_H[ P_{V|X,H}[A] ]$. 
A similar relation holds for $P_{V|\tilde{X}}$.
Also, the divergence measure $D(~||~)$ is jointly convex in its arguments.  
Therefore, by Jensen's inequality,
\begin{eqnarray}
D(~P_{V|X} ~|| ~P_{V|\tilde{X}}~ )  & \leq & E_H[  D(~P_{V|X,H} ~|| ~P_{V|\tilde{X},H}~ )  ] \\  
& = & E_H \left[ || XH-\tilde{X}H ||^2_2 \right]  \\
& = & E_H \left[ \int_0^T |X(t)-\tilde{X}(t)|^2 |H(t)|^2 dt \right] \\
& =  & || X -\tilde{X}||_2^2
\end{eqnarray}
The $L^1$ or variational distance between two probability measures
is bounded by their divergence: namely,  $||P-Q||_1 \leq \sqrt{2D(P||Q)}$
 \cite[Lemma 16.3.1]{CoverThomasDivergence}. So
\begin{equation}
|| P_{V|X}-P_{V|\tilde{X}}||_1  \leq\sqrt{2} ||X-\tilde{X}||_2. 
\end{equation}
 In particular, $P_{V|X}$ as a function of $X$ is a continuous mapping from the space
$L^2[0,\infty)$ to the space of measures with the $L^1$ metric. 
The proof of the lemma is completed by invoking the fact that the divergence function
 $D( P || Q)$ is lower semi-continuous in $(P,Q)$ under the $L^1$ metric 
 (see theorem of Gelfand, Yaglom, and Perez \cite[(2.4.9)]{Pinsker}).
\end{proof}

\begin{lemma} \label{lemma.CPeqCKK}
$C_p(P)=\lim_{K\rightarrow\infty} C_p^{K,K}(P)$
\end{lemma}
\begin{proof} Proposition \ref{prop.peak_equiv} applied to the discrete-time channel that results from the use of class $K$ codes and class $K$ decoders yields:
\begin{equation}    \label{eq.Cp1}
C_p^{K,K}(P)=\sup_{\mbox{$X$ of class $K$}} \frac{ D(P_{Y^K|X}||P_{Y^K|0})  }{||X||^2_2}.
\end{equation}
Lemmas \ref{lemma.cppsup} and \ref{lemma.Dcontinuity},  and the fact that finite class
signals are dense in the space of all square integrable signals implies that
\begin{equation}    \label{eq.Cp2}
C_p(P)= \lim_{K\rightarrow\infty}  \sup_{X \mbox{ of class } K}  \frac{ D(P_{V|X}||P_{V|0})  }{||X||^2_2}.
\end{equation}    
Let ${\cal F}^K$ denote the $\sigma$-algebra generated by the entire observation
process $(Y^K(i): i\geq 0)$, or equivalently, by $(V(2^{-K} i) : i\geq 0)$.  Then
${\cal F}^K$ is increasing in $K$, and the smallest $\sigma$-algebra containing ${\cal F}^K$
for all $K$ is ${\cal F}^V$, the $\sigma$-algebra generated by the observation process $V$.
Therefore, by a property of the divergence measure (see Dobrushin's theorem \cite[(2.4.6)]{Pinsker}),  for any fixed signal $X$,
$D(P_{V|X}||P_{V|0}) = \lim_{K\rightarrow\infty} D(P_{Y^K|X}||P_{Y^K|0}).$
Applying this observation to (\ref{eq.Cp2}) yields
\begin{equation}   \label{eq.Cp3}
C_p(P)= \lim_{K\rightarrow\infty} \sup_{\mbox{$X$ of class $K$}} 
\frac{ D(P_{Y^K|X}||P_{Y^K|0})  }{||X||^2_2}.
\end{equation}
Combining (\ref{eq.Cp1}) and (\ref{eq.Cp3}) yields the lemma. 
\end{proof}

\begin{lemma}  \label{lemma.KKlim}
$ \lim_{K\rightarrow\infty} C^{K,K}_p(P)$ is given by the formula for $C_p(P)$ in Proposition \ref{prop.continuous}.
\end{lemma}

\begin{proof}
Let ${\cal C}$ be a $(T, M,\nu, P, \epsilon)$ code for the continuous time channel with class $K$ codewords. 
Let $T = n 2^{-K}$ for some $n \in {\emph N}$. 
Fix a codeword $$ X(t) = \sqrt{2^{K}} \sum_{i =0}^{n-1} a_i u(2^K t - i)$$ where $u(t) = I_{\{t \in [0,1] \}}$. 

\begin{figure}
\begin{center}
\epsfbox{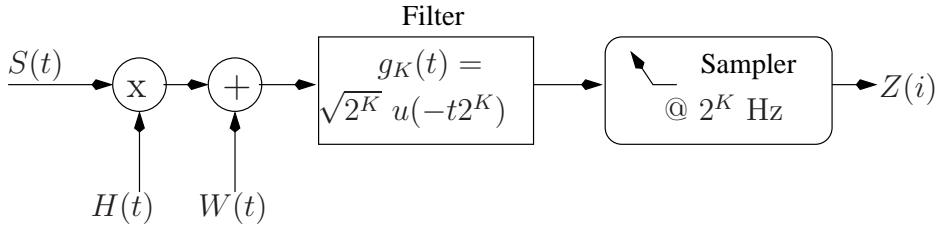} 
\caption{Matched filter response is sampled at rate $2^K$ Hz} \label{fig:ContinuousToDiscreteSystem}
\end{center}
\end{figure}
\begin{figure}
\begin{center}
\epsfbox{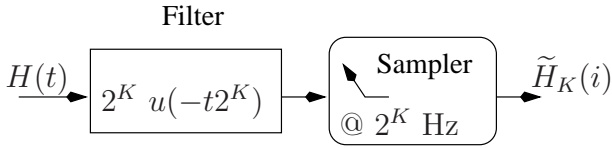} 
\caption{The filter response of the channel process $H(t)$ is sampled ( @ $2^K$ Hz) to generate the discrete time process $\widetilde{H}_K$} \label{fig:MatchedFilterSampler}
\end{center}
\end{figure}
An equivalent discrete time system is constructed using a matched filter at the output, followed by a sampler that generates $2^K$ samples per second, as shown in Figure \ref{fig:ContinuousToDiscreteSystem}. The matched filter $g_K(t)$ is given by 
\begin{equation}
g_K(t) = \sqrt{2^K}u(-t 2^K)
\end{equation}

The equivalent system is 
\begin{equation}
Z(i) = a_i \widetilde{H}_K(i) +\widetilde{W}(i) \label{eq:EquivalentDiscreteTimeSystem}
\end{equation}
Here, the discrete-time process $\widetilde{H}_K$ is a proper complex Gaussian process defined as the filter response of the channel process $H(t)$, sampled at $2^K$ Hz, as shown in Figure \ref{fig:MatchedFilterSampler}. 
The noise process $\widetilde{W}$ is an i.i.d proper complex Gaussian process with zero mean and unit variance.  
The input codeword $X(t)$ in the continuous time system corresponds to an input codeword $\underline{a} = (a_1, \hdots, a_n)$ for the discrete-time system (\ref{eq:EquivalentDiscreteTimeSystem}).

The codebook ${\cal C}$ corresponds to an $(n,M,\nu, P2^{-K}, \epsilon)$ code for the channel $\widetilde{H}_K$. Thus, $C_p^{K,K}(P)$ is the capacity per unit energy $C_p(P 2^{-K})$ of the discrete-time channel process $\widetilde{H}_K$ with peak constraint $P2^{-K}$.

It is easy to see that the spectral density $\{\widetilde{S}_K(\omega): -\pi \leq \omega \leq \pi \}$ of the process $\widetilde{H}_K$ is given by: 
\begin{equation}
\widetilde{S}_K(\omega) = 2^K \sum_{n = -\infty}^{\infty} S \left( 2^K(\omega - 2\pi n) \right) \mbox{sinc}^2(\omega - 2 \pi n)   \label{eq:DiscretizedChannelSpectralDensityK}
\end{equation}
where $$\mbox{sinc}(\omega) \dot{=} \frac{\sin( \omega/2)}{\omega/2}.$$ Let $b_K = R_{\widetilde{H}_K}(0)$ where $R_{\widetilde{H}_K}$ is the autocorrelation function of the process $\widetilde{H}_K$.

\begin{claim} \label{claim.b_KisOne} $\lim_{K\to \infty} b_K$ exists and equals $1$.
\end{claim}
\begin{proof} Clearly $b_K = \int_{-\pi}^{\pi} \widetilde{S}_K(\omega) \frac{d\omega}{2\pi}$. From  (\ref{eq:DiscretizedChannelSpectralDensityK}), it follows that 
\begin{eqnarray*}
b_K &=& \sum_{n = -\infty}^{\infty} \int_{(n-\frac 1 2 )~\pi 2^K}^{(n+\frac1 2)~\pi 2^K} S(\omega) \mbox{sinc}^2 \frac \omega {2^K} \frac {d \omega} {2\pi} \\
 &=& \int_{\omega = -\infty}^{\infty}  S(\omega) \mbox{sinc}^2 \frac \omega {2^K} \frac {d \omega} {2\pi} 
\end{eqnarray*}
Claim \ref{claim.b_KisOne} follows from the Dominated Convergence Theorem, noting that, 
for any $\omega$, \\
$\lim_{K\to \infty} S(\omega) \mbox{sinc}^2 \frac \omega {2^K} = S(\omega)$ and  $\mbox{sinc}^2 \frac \omega {2^K} \leq 1$.
\end{proof}

Let $I_K$ be defined as follows.
\begin{equation}
I_K = 2^K \int_{-\pi}^\pi \log \left( 1+P 2^{-K} \widetilde{S}_K(\omega) \right) \frac {d \omega}{2 \pi} \label{eq:I_KExpression}
\end{equation}
By Claim \ref{claim.b_KisOne}, it follows that 
\begin{equation}
\lim_{K \to \infty} C_p^{K,K}(P) = 1 - \frac 1 P  \lim_{K \to \infty} I_K \label{eq:CapacityPerUnitEnergyContinuousTimeExpression}
\end{equation}

\begin{claim} \label{claim.IntegralCtsTimeExpression} $\lim_{K \to \infty} I_K = \int_{\omega = -\infty}^{\infty} \log \left( 1+P S(\omega) \right) \frac {d \omega} {2\pi}$.
\end{claim}
\begin{proof} 
Substituting for $\widetilde{S}_K$ (\ref{eq:DiscretizedChannelSpectralDensityK}) in the expression for $I_K$ in (\ref{eq:I_KExpression}) yields
\begin{equation*}
I_K = \int_{-\pi 2^K}^{\pi 2^K} \log \left( 1+ P \sum_{n = -\infty}^{\infty} S(\omega - 2\pi n 2^K) \mbox{sinc}^2 (\frac \omega {2^K} - 2 \pi n ) \right) \frac {d \omega}{2 \pi}
\end{equation*}
Fatou's Lemma yields the following lower bound.
\begin{equation}
\liminf_{K > 0} I_K \geq  \int_{\omega = -\infty}^{\infty} \log \left( 1+P S(\omega) \right) \frac {d \omega} {2\pi} \label{eq:IntegralCtsTimeExpressionLowerBound}
\end{equation}
The following upper bound on $I_K$ follows from the fact that for any $x_1 > 0, ~x_2 >0$, $\log(1+x_1+x_2) \leq \log(1+x_1) + x_2$, and $\mbox{sinc}^2(x) \leq 1$ for each $x$:
\begin{eqnarray*}
I_K & \leq & \int_{-\pi 2^K}^{\pi 2^K} \log \left( 1+ P S(\omega ) \mbox{sinc}^2 (\frac \omega {2^K} ) \right) \frac {d \omega}{2 \pi} + \int_{|\omega| > \pi 2^K} S(\omega) \mbox{sinc}^2  (\frac \omega {2^K} )  \frac {d \omega}{2 \pi} \\
 & \leq & \int_{-\pi 2^K}^{\pi 2^K} \log \left( 1+ P S(\omega ) \right) \frac {d \omega}{2 \pi} + \int_{|\omega| > \pi 2^K} S(\omega) \frac {d \omega}{2 \pi}
\end{eqnarray*}
It follows that
\begin{equation}
\limsup_{K>0} I_K \leq \int_{ -\infty}^{\infty} \log \left( 1+P S(\omega) \right) \frac {d \omega} {2\pi} \label{eq:IntegralCtsTimeExpressionUpperBound}
\end{equation}
From (\ref{eq:IntegralCtsTimeExpressionLowerBound}) and (\ref{eq:IntegralCtsTimeExpressionUpperBound}), $\lim_{K\to \infty} I_K$ exists and Claim \ref{claim.IntegralCtsTimeExpression} is proved.
\end{proof}

Claim \ref{claim.IntegralCtsTimeExpression} and (\ref{eq:CapacityPerUnitEnergyContinuousTimeExpression}) complete the proof of Lemma \ref{lemma.KKlim}. 

\end{proof} 

The validity of Proposition \ref{prop.continuous} is implied by Lemmas \ref{lemma.CPeqCKK} and \ref{lemma.KKlim}.  The proof of Proposition \ref{prop.continuous}  is complete.

\section{CONCLUSION} \label{sec.conclusion}
This paper provides a simple expression for the capacity per unit energy of a discrete-time Rayleigh fading channel with a hard  peak constraint on the input signal. 
The fading process is stationary and can be correlated in time. 
There is no channel state information at the transmitter or the receiver.
The capacity per unit energy for the non-coherent channel is shown to be that of the channel with coherence minus a penalty term corresponding to the rate of learning the channel at the output. 
Further, ON-OFF signaling is found to be sufficient for achieving the capacity per unit energy. 
Similar results are obtained for  continuous-time  channels also. 
One application for  capacity per unit energy is to bound from above the capacity per unit time. Upper bounds to capacity per unit time are plotted for channels with Gauss Markov fading.

A possible extension of this paper is to a multiple antenna (MIMO) scenario. While the results may extend in a straightforward fashion to parallel independent channels, extension to more general MIMO channels seems non-trivial.
Also, the fading could be correlated both in time and across antennas. Suitable models of fading channels that abstract such correlation need to be constructed.
Another possible extension of this paper is to consider more general fading models such as the  WSSUS fading model used in \cite{MedardGallager02,SubramanianHajek02}.
This would let us explore the effect of multipath or inter-symbol interference on capacity in the low SNR regime. 
\appendices
\section{BOUNDING CAPACITY PER UNIT ENERGY USING FOURTHEGY} \label{app:FourthegyBoundDerivation}
We bound the capacity per unit energy for the channel in (\ref{eq:ChannelModel}) by applying a bound of M\'edard and Gallager \cite{MedardGallager02}. In the terminology of  \cite{SubramanianHajek02}, this amounts to bounding  the fourthegy using the given average and peak power constraint, and using the expression for capacity per unit fourthegy.

Let $\widetilde{H}$ denote the block fading process such that blocks of length $T$ are independent, with each block having the same probability distribution as $\left(H(1),H(2),\ldots,H(T)\right)$. 
Denote $T$ consecutive uses of a channel with fading process $\widetilde{H}$ by the following:
\begin{eqnarray}
\widetilde{Y}_{T\times 1} & = & \widetilde{H}_{ T\times T} X_{ T\times 1} + W_{T\times 1} \label{eq:MemorylessBlockFadingChannel}\\
E[|X(i)|^2] &\leq & P_{avg} \label{eq:AveragePowerVectorConstraint}\\
|X(i)|^2 & \leq & P ~ \forall ~ i \in \{1 \hdots T\} \label{eq:PeakPowerVectorConstraint}
\end{eqnarray}
Here, $\widetilde{H}_{ T\times T}$ is a diagonal matrix with entries along the main diagonal corresponding to $(\widetilde{H}(1)$,..., $\widetilde{H}(T))$. The average and peak power constraints are specified by (\ref{eq:AveragePowerVectorConstraint}) and (\ref{eq:PeakPowerVectorConstraint}). 
According to a bound of M\'{e}dard and Gallager \cite{MedardGallager02} (also see \cite[Prop. II.1]{SubramanianHajek02}):
\begin{equation}
I(\widetilde{Y}_{T\times 1};X_{T\times 1}) \leq  \frac{1}{2\sigma^4}~E[J_C(X_{T\times 1})], \label{eq:MutualInfo_fourthegy}
\end{equation}
where $J_C(X_{T\times 1})$ is the fourthegy of $\widetilde{Y}_{T \times 1}$ corresponding to input $X_{T\times 1}$.
Normalizing with respect to the additive noise power, $\sigma^2$ is set to $1$.
Let $(Y_{T\times 1};X_{T\times 1})$ denote $T$ consecutive uses of the channel modeled in (\ref{eq:ChannelModel}). 
Since $(Y_{T\times 1};X_{T\times 1})$ and $(\widetilde{Y}_{T\times 1};X_{T\times 1})$ are statistically identical, $I(\widetilde{Y}_{T\times 1};X_{T\times 1})  = I(Y_{T\times 1};X_{T\times 1})$ and the fourthegy of $Y_{T\times 1}$ is also given by $J_C(X_{T\times 1})$.  
The average fourthegy is upper-bounded in the following manner:
\begin{eqnarray*}
J_C(X_{T\times 1}) & = & \sum_{i=1}^T|X_i|^2 \left( \sum_{j=1}^T |X_j|^2 |R_H(i-j)|^2\right) \\
 & \leq & \sum_{i=1}^T |X_i|^2 \left( \sum_{j=1}^T P. |R_H(i-j)|^2\right).
\end{eqnarray*}
The above inequality follows from the peak power constraint (\ref{eq:PeakPowerVectorConstraint}). 
We further upper-bound the above expression and apply Parseval's theorem to obtain 
\begin{eqnarray*}
\sum_{j=1}^T |R_H(i-j)|^2 &\leq& \sum_{j=-\infty}^{\infty} |R_H(i-j)|^2 \\
&=&  \int_{-\pi}^{\pi}S^2(\omega)\,\frac{d\omega}{2\pi}
\end{eqnarray*}
This yields the following upper-bound on fourthegy:
\begin{equation}
J_C(X_{T\times 1}) \leq \sum_{i=1}^T |X_i|^2 \cdot P \int_{-\pi}^{\pi}S^2(\omega)\,\frac{d\omega}{2\pi} \label{eq.FourthegyUpperBound}
\end{equation}
Combining (\ref{eq.FourthegyUpperBound}) and (\ref{eq:MutualInfo_fourthegy}) yields 
\begin{eqnarray}
\frac{I(Y_{T\times 1};X_{T\times 1})}{E\left[\sum_{i=1}^T |X_i|^2 \right]} & \leq &  U_p \label{eq:FourthegyUpperBoundOnCapPerUnitEnergy}
\end{eqnarray}
where $U_p$ is given in Equation (\ref{eq:FourthegyBound}).
From (\ref{eq:FourthegyUpperBoundOnCapPerUnitEnergy}) and (\ref{eq:CapEnergyMutualInfoConnectVerdu}), it follows that $C_p \leq U_p$. 

\section{PROOF OF COROLLARY \ref{cor.GaussMarkov}} \label{app.GaussMarkovProof}
The fading process $H$ is Gauss Markov with autocorrelation function $\rho^{|t|}$ for some
$\rho$ with $0 \leq  \rho < 1$. By Proposition \ref{prop.main}, the capacity per  unit energy for peak constraint $P$ is given by (\ref{eq:C5}). The expression $ \frac{1}{P} \int_{-\pi}^{\pi} \log(1+P S(\omega))\,\frac{d\omega}{2\pi}$ is now evaluated for the Gauss Markov fading process.
The autocorrelation function $R_H$ of the  Gauss Markov fading process is given by $R_H(n) = \rho^{|n|}$. Its z-transform, $S(z)$ is given by
\begin{equation*}
S(z) = \frac{1-\rho^2}{(1-\rho z)(1-\rho z^{-1})} \mbox{ for } |z|<1.
\end{equation*}
Note that $1+P S(z)$ is a rational function with both numerator and denominator having degree two. Zeros of the function $1+P S(z)$ satisfy 
\begin{equation}
 \rho^2 z^2 + \rho z \{1+P+\rho^2(1-P)\} + \rho^2 = 0. \label{eq:roots_numerator}
\end{equation} 
Recall that $z_+$ is the larger root of the equation
\begin{equation}
z^2 + z \{1+P+\rho^2(1-P)\} + \rho^2 = 0.
\end{equation}
Comparing the two equations, it follows that $\frac{z_+}{\rho}$ is a zero of $1+P S(z)$. Since $R_H(n)$ is even, $S(z) = S(z^{-1})$. So, the other zero of $1+P S(z)$ is $\frac{\rho}{z_+}$. (This is also evident since the product of the roots of (\ref{eq:roots_numerator}) is $1$.) It follows that $1+P S(z)$ can be written as 
\begin{eqnarray*}
1+P S(z) &=& \frac{ -\rho z +\left\{1+P+\rho^2(1-P)\right\}-\rho z^{-1} }{(1-\rho z)(1-\rho z^{-1}) }\\
 	&=& \frac{ (-1/\rho z) \left\{\rho^2 (z-\frac{z_+}{\rho})(z-\frac{\rho}{z_+})\right\} }{ \rho^2  (z-\frac{1}{\rho})(z^{-1}-\frac{1}{\rho} ) }
\end{eqnarray*}
Consider the terms in the numerator of the above expression. Since $(-1/\rho z)(z-\frac{\rho}{z_+})= (1/z_+)(z^{-1} - \frac{z_+}{\rho})$, $1+P S(z)$ can be further simplified as 
\begin{equation}
1+P S(z) = \frac{ (z-\frac{z_+}{\rho})(z^{-1} - \frac{z_+}{\rho}) }{ z_+ (z-\frac{1}{\rho})(z^{-1}-\frac{1}{\rho} ) }.
\end{equation}
Hence for $|z|=1$,
\begin{equation}
1+PS(z) = |f(z)| \mbox{ where } f(z) = \frac{ (z-\frac{z_+}{\rho})^2 }{ z_+(z-\frac{1}{\rho})^2 }.
\end{equation}
Since the polynomial in (\ref{eq:roots_numerator}) is negative at $z=1$ and positive as $|z| \to \infty$, it is clear  that $z_+ > 1 $. The function $f$ is analytic and nonzero in a neighborhood of the unit disk. Thus, by Jensen's formula of complex analysis,
\begin{eqnarray*}
 \int_{-\pi}^{\pi} \log(1+P S(\omega))\,\frac{d\omega}{2\pi} &=&\int_{-\pi}^{\pi} \log|f(e^{j\omega})|\,\frac{d\omega}{2\pi} \\
 &=& \log|f(0)|\\
 &=& \log z_+.
\end{eqnarray*}
Equation (\ref{eq:EnergyBoundGaussMarkov}) in Corollary \ref{cor.GaussMarkov} follows.

The integral in the expression for $U_p$, as given in (\ref{eq:FourthegyBound}) is simplified using Parseval's theorem as follows:
\begin{equation*}
\int_{-\pi}^{\pi}S^2(\omega)\,\frac{d\omega}{2\pi} = \sum_{-\infty}^{\infty} \rho^{2|n|} = \frac{1+\rho^2}{1-\rho^2} .
\end{equation*}
Equation (\ref{eq:FourthegyBoundGaussMarkov})  in Corollary \ref{cor.GaussMarkov} follows.
\section{PROOF FOR COROLLARY \ref{cor.BlockFadingDiscreteAndCts}} \label{app:BlockFadingProof}
The proof works for both discrete-time and continuous-time channels.
Let $\Gamma$ be the input alphabet ($\Gamma= {\cal C}^T$ for a discrete-time channel).
Since the block fading channel is a discrete memoryless vector channel, Verd\'u's formulation \cite{Verdu90} of capacity per unit cost applies here. 
\begin{equation}
C_p^B(P,T) = \sup_{X \in \Gamma: ~X \neq 0} \frac{D(p_{Y|X}\|p_{Y|0})}{\|X\|_2^2},
\end{equation}
where $X$ is understood to satisfy the peak power constraint $\|X\|_{\infty} \leq \sqrt{P}$. 
Following \cite[p. 812]{SubramanianHajek02} (discrete-time) and \cite[Prop. III.2, (16)]{SubramanianHajek02} (continuous-time), $D(p_{Y|X}\|p_{Y|0})$ can be expressed as $\sum_i \phi(\lambda_i)$, where $\phi(\lambda) = \lambda - \log(1+\lambda)$ and $\{\lambda_i\}$ is the set of  eigenvalues of the autocorrelation matrix (discrete-time) or autocorrelation function (continuous-time) of the signal $HX$. This signal has rank one. So, $\lambda_1 = \|X\|_2^2$, and $\lambda_i=0$ for $i\neq 1$. Thus, 
\begin{equation}
D(p_{Y|X}\|p_{Y|0}) = \|X\|_2^2 - \log(1+\|X\|_2^2).
\end{equation}
So, the expression for capacity per unit energy with peak constraint $P$ simplifies to:
\begin{equation}
C_p^B(P,T) = 1 - \inf_{X \in \Gamma: ~\|X\|_{\infty}^2 \leq P} \frac{\log (1+ \|X\|_2^2)}{\|X\|_2^2}.
\end{equation}
Since $\frac{log(1+x)} x$ is monotonic decreasing in $x$ for $x>0$, the above infimum is achieved when $\|X\|_2^2$ is set at its maximum allowed value $PT$. This completes the proof of Corollary \ref{cor.BlockFadingDiscreteAndCts}. 

\section{ALTERNATIVE PROOF FOR LEMMA \ref{lem:LogDetLemma}} \label{app:LogDetLemmaAlternateProof}
Lemma \ref{lem:LogDetLemma} shows that, in the limit as $n \to \infty$, the expression \[ \frac{1}{n} \log \det(I_n + P\cdot \Sigma_n)\]
can be expressed as an integral involving $S(\omega)$, the density of the absolutely continuous part of the spectral measure of the fading process $H$. Here, we present a brief outline on an alternative proof for the same. 

The term $\det(I+P \Sigma_n)$ can be expanded as a product of its eigenvalues. 
\begin{equation}
\frac{\log \det(I+P\Sigma_n)}{nP} = \frac{1}{P} \left[\frac{1}{n}\sum_{i=1}^{n}\log(1+P \lambda_i) \right] 
\end{equation}
Here, $\lambda_i$ is the $i^{th}$ eigenvalue of $\Sigma_{n\times n}$.
The  theory of circulant matrices is now applied to evaluate this limit as an integral:
\begin{equation}
\lim_{n\to \infty} \frac{1}{n}\sum_{i=1}^{n}\log(1+P \lambda_i)  = \int_{-\pi}^{\pi} \log(1+P S(\omega))\,\frac{d\omega}{2\pi}.
\end{equation}
The result about convergence of the log determinants of Toeplitz matrices is known as Szeg\"{o}'s first limit theorem and was established by Szeg\"{o} \cite{Szego}.  Later it came to be used in the theory of linear prediction of  random processes (see for example \cite[\S ~IV.9 Theorem 4]{GihmanSkorohod}). Therefore,
\begin{equation}
\lim_{n\to \infty} \frac{\log \det(I+P\Sigma_n)}{nP}  = \frac{1}{P} \int_{-\pi}^{\pi} \log(1+P S(\omega))\,\frac{d\omega}{2\pi}.
\end{equation}
\section{PROOF OF LEMMA \ref{lem.CapacityPerUnitEnergyOfBlockChannelWithCSI} } \label{app.CapacityPerUnitEnergyOfBlockChannelWithCSIProof}
Since the channel modeled in (\ref{eq:BlockFadingCoherenceChannel}) is a discrete-time memoryless vector channel, the formulation of capacity per unit cost in \cite{Verdu90} can be applied. 
\begin{equation}
C_{p,T}^{coh}(P)  =  \sup_{X:\|X\|_{\infty}\leq {\sqrt P}} \frac{D(p_{\hat{Y}|X}\|p_{\hat{Y}|0})}{\|X\|_2^2} \label{eq:CapacityOfBlockChannelWithCSIDivergenceExpression}
\end{equation} 
Here, $X$ is a deterministic complex vector in $C^T$.   
Let $\Sigma_X$ denote the covariance matrix of $\hat{Y}$ conditional on $X$ being transmitted, and $\Sigma_0$ the covariance matrix of $\hat{Y}$ conditional on $0$ being transmitted.
Let $\bar{X}$ denote $diag(X)$, and $\Sigma$ denote the covariance matrix of the random vector $(H(1), \ldots, H(T))^T$. The following expressions for $\Sigma_X$ and $\Sigma_0$ are immediate. 
\begin{eqnarray}
\Sigma_X &=& \left( \begin{array}{cc}
\bar{X}\Sigma \bar{X}^{\dag}+I_T & \bar{X} \Sigma \\
\Sigma \bar{X}^{\dag}            & \Sigma \end{array} \right) \\
\Sigma_0 &=& \left( \begin{array}{cc}
I_T & 0\\
0   & \Sigma \end{array} \right)
\end{eqnarray}
The divergence expression in (\ref{eq:CapacityOfBlockChannelWithCSIDivergenceExpression}) then simplifies to 
\begin{eqnarray}
D(p_{\hat{Y}|X}\|p_{\hat{Y}|0}) &=& \log \frac{\det \Sigma_X}{\det \Sigma_0} + tr \left(E_{p_{\hat{Y}|X}}[\Sigma_0^{-1}\hat{Y}\hat{Y}^\dag - \Sigma_X^{-1}\hat{Y}\hat{Y}^\dag]\right) \\
&=& \log \frac{\det \Sigma_X}{\det \Sigma_0} + tr \left( \Sigma_0^{-1} \Sigma_X - I_{2T} \right) \\
&=&  \log \frac{\det \Sigma_X}{\det \Sigma_0} + tr \left( \begin{array}{cc} \bar{X}\Sigma \bar{X}^{\dag} & 0 \\
0 & 0 \end{array} \right) \label{eq:CSIChannelDivergenceExpressionStep1}
\end{eqnarray}
It is clear that $\det \Sigma_0 = \det \Sigma$. To evaluate $\det \Sigma_X$, let $\hat \Sigma_X$ be given by
\begin{equation*}
\hat \Sigma_X = \left( \begin{array}{cc} Re(\Sigma_X) & -Im(\Sigma_X) \\
Im(\Sigma_X) &  Re(\Sigma_X) \end{array} \right).
\end{equation*}
Clearly, $\det \hat \Sigma_X  = \left( \det \Sigma_X \right)^2 $. Since $\hat \Sigma_X$ is a matrix with real entries, row-operations leave the determinant unchanged yielding the following expression.
\begin{eqnarray*}
\det \hat \Sigma_X &=& \det \left[ \left( \begin{array}{cc} I & 0 \\ \Sigma \bar{X}^{\dag} & \Sigma \end{array} \right)
 \left( \begin{array}{cc} I & \bar{X} \Sigma \\ 0 & \Sigma \end{array} \right) \right] \\
  &=& (\det \Sigma)^2. \\
\mbox{So, } \det \Sigma_X &=& \det \Sigma. 
\end{eqnarray*}
This implies that $\log \frac{\det \Sigma_X}{\det \Sigma_0} = 0$. Since the correlation matrix $\Sigma$ is normalized to have ones on the main diagonal, $tr (\bar{X}\Sigma\bar{X}^{\dag}) = \sum_{i=1}^T |X_i|^2$. So, the divergence expression in (\ref{eq:CSIChannelDivergenceExpressionStep1}) evaluates to $1$ independent of the choice of the deterministic complex vector $X$ (as long as $X \neq 0$). This, along with (\ref{eq:CapacityOfBlockChannelWithCSIDivergenceExpression}), proves (\ref{eq:CapacityPerUnitEnergyOfBlockChannelWithCSI}).

{\bf Vignesh Sethuraman} (S'04) received his B.Tech. from Indian Institute of Technology, Madras, in 2001, and M.S. from University of Illinois at Urbana-Champaign in 2003. He is currently with working towards the Ph.D. degree. His research interests include communication systems, wireless communication, networks and information theory.

{\bf Bruce Hajek} (M'79-SM'84-F'89) received a B.S. in Mathematics
and an M.S. in Electrical Engineering from the University
of Illinois in 1976 and 1977, and a Ph. D. in Electrical
Engineering from the University of California at Berkeley in 1979.
He is a Professor in the Department of Electrical and Computer
Engineering and in the Coordinated Science Laboratory at the
University of Illinois at Urbana-Champaign, where he has been
since 1979.  He served as Associate Editor for Communication
Networks and Computer Networks for the IEEE Transactions
on Information Theory (1985-1988), as Editor-in-Chief  of the same
Transactions (1989-1992), and as President of the IEEE Information
Theory Society (1995).  His research interests include communication
and computer networks, stochastic systems, combinatorial and nonlinear
optimization and information theory. Dr. Hajek is a Member of the National
Academy of Engineering and he received the IEEE Koji Kobayashi Computers and
Communications Award, 2003.

\renewcommand{\theequation}{A-\arabic{equation}}
\setcounter{equation}{0} 


\begin{thebibliography}{10}

\bibitem{Verdu90}
S.~Ver{d\'{u}}, ``On channel capacity per unit cost,'' {\em IEEE Transactions
  on Information Theory}, vol.~36, pp.~1019--1030, Sept. 1990.

\bibitem{Gallager87}
R.~G. Gallager, ``Energy limited channels: Coding, multiaccess, and spread
  spectrum,'' {\em {Tech. Report, LIDS-P-1714}}, Nov. 1987.

\bibitem{AbouFaycalTrottShamai01}
I.~C. Abou-Faycal, M.~Trott, and S.~Shamai, ``The capacity of discrete-time
  memoryless {R}ayleigh fading channels,'' {\em IEEE Transactions on
  Information Theory}, vol.~47, pp.~1290--1301, May 2001.

\bibitem{MedardGallager02}
{M.~M\'{e}dard and R. G. Gallager}, ``Bandwidth scaling for fading multipath
  channels,'' {\em IEEE Transactions on Information Theory}, vol.~48,
  pp.~840--852, Apr. 2002.

\bibitem{SubramanianHajek02}
V.~G. Subramanian and B.~Hajek, ``Broad-band fading channels: Signal burstiness
  and capacity,'' {\em IEEE Transactions on Information Theory}, vol.~48,
  pp.~809--827, Apr. 2002.

\bibitem{Hassibi04}
C.~Rao and B.~Hassibi, ``Analysis of multiple-antenna wireless links at low
  {SNR},'' {\em IEEE Transactions on Information Theory}, vol.~50,
  pp.~2123--2130, Sept. 2004.

\bibitem{Telatar99}
E.~Telatar, ``Capacity of multi-antenna {G}aussian channels,'' {\em European
  Transactions on Telecommunications}, vol.~10, pp.~585--595, Dec. 1999.

\bibitem{MarzettaHochwald99}
{T.~Marzetta and B.~Hochwald}, ``Capacity of a mobile multiple-antenna
  communication link in {R}ayleigh flat fading,'' {\em IEEE Transactions of
  Information Theory}, vol.~45, pp.~139--158, Jan. 1999.

\bibitem{Lapidoth05}
A.~Lapidoth, ``On the asymptotic capacity of stationary {Gaussian} fading
  channels,'' {\em IEEE Transactions on Information Theory}, vol.~51,
  pp.~437--446, Feb. 2005.

\bibitem{Verdu02}
A.~M. Tulino, A.~Lozano, and S.~Verd{\'{u}}, ``Capacity of multi-antenna
  channels in the low-power regime,'' {\em Information Theory Workshop},
  pp.~192--195, Oct. 2002.

\bibitem{HajekSubramanian02}
B.~Hajek and V.~Subramanian, ``Capacity and reliability function for small peak
  signal constraints,'' {\em IEEE Trans. on Information Theory}, vol.~48,
  pp.~829--839, 2002.

\bibitem{Shannon49}
C.~E. Shannon, ``A mathematical theory of communication,'' {\em Bell Syst.
  Technical Journal}, vol.~27, pp.~379--423, 623--656, July--Oct 1948.

\bibitem{Maruyama49}
G.~Maruyama, ``The harmonic analysis of stationary stochastic processes,'' {\em
  Memoirs of the faculty of science, {S}eries {A}, {M}athematics}, vol.~4,
  no.~1, pp.~45--106, 1949.

\bibitem{Pinsker}
M.~S. Pinsker, {\em Information and Information Stability of Random Variables
  and Processes}.
\newblock Holden-Day, Inc., 1964.

\bibitem{Gray}
R.~M. Gray, {\em Entropy and Information Theory}.
\newblock Springer-Verlag, 1990.

\bibitem{Biglieri98}
E.~Biglieri, J.~Proakis, and S.~Shamai, ``Fading channels:
  Information-theoretic and communication aspects,'' {\em IEEE Transactions on
  Information Theory}, vol.~44, pp.~2619--2692, Oct. 2002.

\bibitem{Jacobs63}
I.~Jacobs, ``The asymptotic behaviour of incoherent {M}-ary communication
  systems,'' {\em Proc. IEEE}, vol.~51, pp.~251--252, Jan. 1963.

\bibitem{Kennedy69}
R.~S. Kennedy, {\em Fading Dispersive Communication Channels}.
\newblock New York: Wiley-Interscience, 1969.

\bibitem{TelatarTse00}
E.~Telatar and D.~Tse, ``Capacity and mutual information of wide-band multipath
  fading channels,'' {\em IEEE Transactions on Information Theory}, vol.~46,
  pp.~2315--2328, Nov. 2000.

\bibitem{TableOfIntegralsGradshteyn}
I.~S. Gradshteyn and I.~M. Ryzhik, {\em Table of Integrals, Series and
  Products}.
\newblock Academic Press, INC., 1980.

\bibitem{Stuber}
G.~L. St\"{u}ber, {\em Principles of Mobile Communication}.
\newblock Kluwer Academic Publishers, 1996.

\bibitem{GihmanSkorohod}
I.~Gihman and A.~Skorohod, {\em The Theory of Stochastic Processes I}.
\newblock New York: Springer-Verlab, 1974.

\bibitem{Doob53}
J.~Doob, {\em Stochastic Processes}.
\newblock New York: Wiley, 1953.

\bibitem{Gallager68}
R.~G. Gallager, {\em Infomation Theory and Reliable Communication}.
\newblock Wiley, 1968.

\bibitem{Wyner88}
A.~D. Wyner, ``Capacity and error exponent for the direct detection photon
  channel, {I--II},'' {\em IEEE Transactions on Information Theory}, vol.~34,
  pp.~1449--1461, 1462--1471, Dec. 1988.

\bibitem{CoverThomasDivergence}
T.~M. Cover and J.~A. Thomas, {\em Elements of Information Theory}.
\newblock Wiley series in telecommunications, 1991.

\bibitem{Szego}
G.~Szeg{\"{o}}, ``Ein grenzwertsatz uber die toeplitzschen determinanten einer
  reelen psitiven funktion,'' {\em Math, Ann.}, vol.~76, pp.~490--503, 1915.

\end{thebibliography}
\end{document}